\documentclass[10pt,aps,prb,twocolumn,amsmath,amssymb,superscriptaddress,floatfix]{revtex4-2}
\usepackage{amsfonts,bm,graphicx} 
\usepackage{braket}
\usepackage{multirow}
\usepackage{hyperref} 
\hypersetup{colorlinks=true,    linkcolor=blue,    citecolor=blue,     urlcolor=blue}
\newcommand{\mk}[1]{\mathbf{#1}}
\newcommand{\overbar}[1]{\mkern 3.5mu\overline{\mkern-3.5mu#1\mkern-1mu}\mkern 1mu}

\begin{document}
\title{Longitudinal magnons in large-$S$ easy-axis magnets}

\author{A. El Mendili}
\affiliation{Universit\'e Grenoble Alpes, CEA, IRIG, PHELIQS, 38000 Grenoble, France}
\author{T. Ziman}
\affiliation{Institut Laue Langevin, 38042 Grenoble Cedex 9, France}
\affiliation{Kavli Institute for Theoretical Science, University of the Chinese Academy of Sciences, Beijing 100190, China}
\author{M. E. Zhitomirsky}
\affiliation{Universit\'e Grenoble Alpes, CEA, IRIG, PHELIQS, 38000 Grenoble, France}
\affiliation{Institut Laue Langevin, 38042 Grenoble Cedex 9, France}

\begin{abstract}
Longitudinal  magnons are a distinct type of multipolar excitations in magnetic materials 
with large spins $S\ge 1$ and strong easy-axis anisotropy. These excitations have  angular momentum 
$S^z = \pm 2S$ and can be viewed as  a propagating full spin reversal.
We study longitudinal magnons for the neareast-neighbor Heisenberg ferromagnet and antiferromagnet
on a square lattice with large single-ion anisotropy.
In the strong-coupling limit,  we derive an effective spin-1/2 model including two leading 
contributions in $J/D$. The effective model provides a simple description of the longitudinal magnon dynamics.
For $S=1$, we compare results from several theoretical approaches that include the effective spin-1/2 model,
the linked-cluster expansion, the multiboson spin-wave theory, and, for a ferromagnet,  an exact two-particle solution.
Among these approaches, the multiboson spin-wave theory provides the decay rate of longitudinal magnons and describes evolution of the excitation spectra from strong to weak anisotropy.
\end{abstract}
\date{\today}

\maketitle

\section{Introduction}
\label{Intro}

Magnetic anisotropy in conjunction with low dimensionality or geometrical frustration can produce fractional  excitations. A well-known example of such process is realized for the $XXZ$ spin-1/2 antiferromagnetic chain, which features a dissociation of one magnon with $|S^z|=1$ into two propagating domain walls or spinons each with $|S^z|=1/2$  \cite{Giamarchi, Mikeska}. 
The spin-ice pyrochlores provide another recent example, wherein a microscopic magnetic dipole splits into two effective magnetic charges (monopoles) \cite{Castelnovo12}.  Even more exotic Majorana fermions can appear in the Kitaev honeycomb-lattice model with anisotropic bond-dependent interactions \cite{Kitaev03}. 

The opposite of fractionalization is the fusion of several integer quasiparticles into a composite excitation. 
The formation of bound magnon pairs in ferromagnets can be considered as an  example of such  a process 
\cite{Wortis63,Hanus63,Torrance1969IBF,Wada75,Mattis86}. In the current literature, the composite magnetic 
excitations are usually referred to as multimagnon bound states 
\cite{Torrance69,Tonegawa71,Silberglitt70,Allen71,Oguchi73,Liu78,Fert78,Psaltakis84,Katsumata00,Kecke2007,Oitmaa08,Fukuhara13,Dally20,Legros2021,Wyzula22,Mardele24, Bai2021,Mook23, Bai23, Delgado24, Sheng25}. 
However, just as there are different classes of fractional magnetic excitations, one can also distinguish several types 
of composite quasiparticles. 

In the present study  we focus on a particular type of composite excitations---the longitudinal magnons. 
In contrast to ordinary magnons, these longitudinal spin modes possess a large angular momentum $| S^z|=2S$. 
Their appearance in collinear ferro- and antiferromagnets with $S\geq 1$ is driven by a large single-ion  anisotropy 
of the easy-axis type:
\begin{equation}
\hat{\cal H} = \hat{\cal H}_{\rm ex}  -D\sum_i (\hat{S}_i^z)^2 \,,
\label{H0}
\end{equation}
where $\hat{\cal H}_{\rm ex}$ describes the exchange interactions. The spin model (\ref{H0}) features a continuous
rotation symmetry about the $z$ axis, which leads to conservation of the $z$ component of the total spin. 
As a result, for zero or weak anisotropy the low-energy excitation spectrum consists of conventional magnons 
with integer quantum spin numbers $S^z= -1$  for ferromagnets  and $S^z= \pm 1$  for antiferromagnets
\cite{Kittel,Auerbach}. On the other hand, in the opposite limit  $D/|J|\gg 1$ where $J$ is the typical exchange constant,
the lowest-energy excitations are spin reversals: $\ket{\pm S}\rightarrow\ket{\mp S}$.  Given their quantum spin 
number $S^z=\pm 2S$, the longitudinal magnons can also be described  as  bound states of $2S$ magnons.

The theoretical prediction of new magnetic excitations in the easy-axis ferromagnets, induced by 
the single-ion anisotropy, dates back to the 1970s. Besides exchange bound states, a spin-$S$ 
ferromagnet (\ref{H0}) exhibits an additional branch of two-magnon bound states called single-ion 
bound states \cite{Tonegawa71,Silberglitt70,Liu78}. The term `two-magnon single-ion bound states' has 
been carried over to the experimental studies, which so far have mostly focused on the spin-1 
magnetic materials \cite{Fert78,Psaltakis84,Katsumata00,Legros2021,Bai2021,Sheng25}.

\begin{figure}[tb]
\centering
\includegraphics[width=0.99\columnwidth]{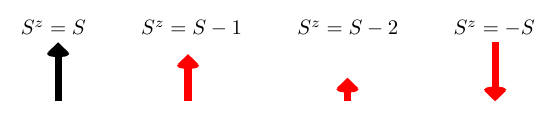}
\includegraphics[width=0.305\columnwidth]{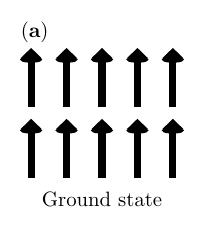}
\label{subf_scheme_gs}
\includegraphics[width=0.3\columnwidth]{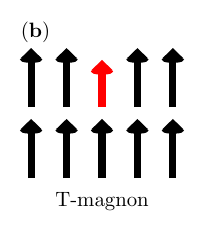}
\label{subf_scheme_TM}
\includegraphics[width=0.3\columnwidth]{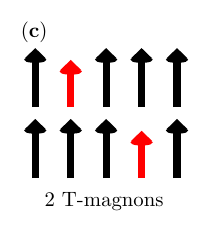}
\label{subf_scheme_2uTM}
\vspace*{-0.25em}
\includegraphics[width=0.32\columnwidth]{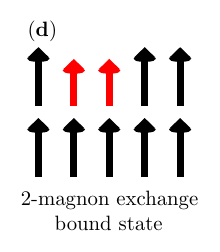}
\label{subf_scheme_2bTM}
\hspace{-0.4em}
\includegraphics[width=0.325\columnwidth]{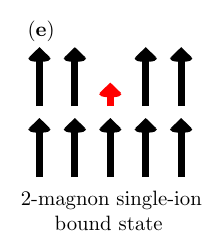}
\label{subf_scheme_2SIBS}
\hspace{-0.1em}
\includegraphics[width=0.297\columnwidth]{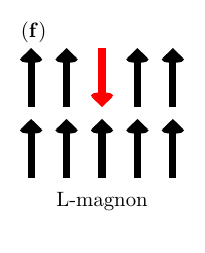}
\label{subf_scheme_LM}
\vspace*{-0.25em}   
\caption{Schematic representation of the magnetic structure (a) and low-lying excitations (b-f) in the spin-$S$ 
easy-axis ferromagnet. The top row shows a chart  for different  quantum states of a single spin.}
\label{fig_scheme}
\end{figure}

Note that for $S=1$ the two magnon bound states are the only possible type of single-ion excitations. 
For larger spins, $S>1$, the Hilbert space expands, allowing for several kinds of single-ion bound states. 
Their classification can be straightforwardly developed starting with the large-$D$ limit.  A schematic picture 
of the low-energy excitations is shown in Fig.~\ref{fig_scheme}. For $D\gg |J|$, one-magnon excitations with 
$| S^z| = 1$ have energy $E_1 \simeq D(2S-1)$, where we keep only the largest anisotropy term contribution, 
and omit the dispersive part $\sim O(JS)$. Within the same accuracy, the two magnon states in the continuum 
are of energy  $E_{2}\simeq 2D(2S-1)$. A pair of magnons attracted by the exchange interactions lower their 
energy by $\sim O(J)$ compared to the states in the continuum but leave the anisotropic part intact: 
$E_2^{\rm EB}\simeq 2D(2S-1)$.  In addition, there are three families of single-ion  bound-states: 

(i) Two magnons on the same site,  Fig.~\ref{fig_scheme}(e), have  $E_2^{\rm SI}\simeq 4D(S-1)$, 
which is  lower than $E_2$ by $2D$. Therefore, the easy-axis anisotropy promotes creation of bound magnon 
pairs with $|S^z|=2$. 

(ii) A full spin flip or $2S$-magnon bound state, Fig.~\ref{fig_scheme}(f), costs no magnetic anisotropy (\ref{H0}) 
and acquires energy  solely due to the broken exchange bonds: $E_{2S}^{\rm SI}= O(JS^2)$.  These are  
the longitudinal magnons.

(iii) An $n-$magnon single-ion bound state with $| S^z|=n$ for intermediate $2<n<2S$ can also appear. 
Its energy is  $E_n^{\rm SI}\simeq D(2S-n)n$, which is lower  than the energy $E_n$ of $n$ independent 
magnon states in the  continuum. These  states are not included to Fig.~\ref{fig_scheme}.

For $S=1$, the single-ion bound states of  types I and II are the same. For $S=3/2$, there are no states of 
the third type. However, for $S\geq 2$ all three types of single-ion bound states can occur.
The single-ion bound states can equally appear in ferro- and antiferromagnets, whereas the exchange binding 
of magnons relies on the sign of the exchange interactions \cite{Wortis63,Hanus63,Oguchi73}.

The two-magnon single-ion bound states have been  observed experimentally in FeI$_2$ \cite{Fert78,Katsumata00,Legros2021,Bai2021}, FeBr$_2$ \cite{Psaltakis84} and Na$_2$BaNi(PO$_4$)$_2$ \cite{Sheng25} with effective spins $S=1$, and in CoF$_2$ with $S=3/2$ 
\cite{Allen71}. The recent discovery of  $|S^z|=4$ excitations in the  spin-2 antiferromagnets FePS$_3$ \cite{Wyzula22} 
and FePSe$_3$ \cite{Mardele24} gives a clear evidence of the type-II single-ion bound states in the large-$S$ magnetic materials.
A signature of the 3-magnon resonances, type-III states, has been also found for FePSe$_3$ \cite{Mardele24}.
 
For $D\gg |J|$ and arbitrary $S\geq 1$, the type-II bound states become the lowest energy excitations across 
all spin subsectors.
This fact, together with similarity in their theoretical description, allow us to group
these excitations with a varying multipolar moment $|S^z|=2S$ into a separate class of magnetic
quasiparticles in strongly anisotropic collinear magnets. 
Their name stems from an observation that irrespective of the value of their angular moment $S^z$, the type-II single-ion bound states  contribute
to the longitudinal component of the dynamical structure factor $S^{zz}({\bf q},\omega)$, see Sec.~\ref{Sec-S-model}.
To clearly distinguish the longitudinal or $L$-magnons  from ordinary (transverse) magnons the latter 
are labeled as $T$-magnons in the following.

In this article we describe and compare several analytical approaches that  can be applied to study  
 longitudinal magnons.  Calculations are performed for a simple spin model,  the nearest-neighbor Heisenberg 
Hamiltonian with the single-ion term (\ref{H0}), which allows for a comparison of the relative
accuracy of each method. In addition, we determine a range of microscopic parameters, where $L$-magnons 
exist as well-defined excitations. Using the multiboson representation of spins, we also compute the lifetimes of these excitations for $S=1$, thus describing the evolution of their stability from strong to weak anisotropy.

The paper is organized as follows. In Sec.~\ref{STM} we introduce the spin model and summarize 
basic properties of the conventional magnons for both signs of the exchange constant $J$.
Section~\ref{SLM} is devoted to the effective Hamiltonian approach, which is valid in the strong-coupling
limit $D/J\gg 1$ and consists of mapping a spin-$S$ model
on interacting pseudo-spins $s=1/2$. Two leading nontrivial contributions in $J/D$ 
are computed for bond terms in the $XXZ$ spin-1/2 Hamiltonian starting with the original
model for spin values $S=1$, 3/2, 2, and 5/2. These results remain valid  beyond the original
model and can be used for any spin Hamiltonian with 
Heisenberg exchange interactions.
Subsequent calculations in the framework of the multiboson spin-wave theory
are presented separately for  the $S=1$  antiferromagnet in Sec.~\ref{AFM} and  for the $S=1$ ferromagnet in Sec.~\ref{FM}.   In addition, in Sec.~\ref{fm-ex} we revisit the problem of two-magnon bound states in 
ferromagnets with the single-ion anisotropy and present a simple framework for their computation.
Section~\ref{Concl}  provides a summary of the results and further discussion.

\section{Spin-$S$ Model}
\label{Sec-S-model}

In our theoretical study we consider the simplest  spin model with coherent multipolar excitations,
the Heisenberg model on a square lattice with a single-ion term:
\begin{equation}
\hat{\cal H} = J \sum_{\langle ij\rangle} \mathbf{S}_i\cdot \mathbf{S}_j - D \sum_i (S_i^z)^2  \ .
\label{H}
\end{equation}
The nearest-neighbor exchange coupling can be  either ferromagnetic, $J<0$, or antiferromagnetic, $J>0$.
The single-ion constant $D>0$ represents an easy-axis magnetic anisotropy. Accordingly,
 the ordered magnetic structures are simple collinear ferro- and antiferromagnetic states  
with  spins parallel to the $z$ axis. In this section we present analytical results for the low-energy excitations 
for both states and an arbitrary value of spin
 $S\geq 1$.

\subsection{Transverse magnons}
\label{STM}

\subsubsection{Antiferromagnet}

The semiclassical N\'eel state with $\langle S_i^z\rangle = \pm S$  preserves the $U(1)$ rotation 
symmetry of the spin Hamiltonian and is an eigenstate of the total spin with $S^z_{\rm tot}=0$. 
Quantum fluctuations generated by the transverse components of the Heisenberg exchange
reduce the ordered spin moments $\langle |S_i^z|\rangle < S$, but conserve the  spin quantum number 
for the ground state \cite{Auerbach}. Consequently,  $S^z$  remains a good quantum number for  
magnetic excitations. In particular, conventional transverse magnons have $S^z=\pm 1$. Note that
the spin conservation affects their dynamics by
forbidding spontaneous two-magnon decays \cite{MZH13}.

The nonfrustrated geometry of a square lattice leads to moderate quantum effects  in the antiferromagnetic
ground state for $S>1/2$ even in the isotropic limit \cite{Manousakis91}. Zero-point fluctuations are further 
suppressed by a gap that opens  for finite $D/J$. Harmonic spin-wave theory accurately describes
the conventional (transverse) magnons of the spin model (\ref{H}).  Spin-wave calculations are performed 
by representing spins in terms of boson operators via the Holstein-Primakoff transformation.  
The quadratic boson form is diagonalized by applying the Fourier transformation followed by 
the canonical Bogolyubov transformation \cite{Kittel,Mattis}. 
The magnon dispersion is expressed as
\begin{equation}
\varepsilon^T_{\bf k}=4JS\sqrt{\Bigl[1+ \overbar{D}/(2J)\Bigr]^2-\gamma_{\bf k}^2} \ .
\label{et_swt_afm}
\end{equation}
Here, $\gamma_{\bf k}=\frac{1}{2}(\cos{k_x}+\cos{k_y})$ and $\overbar{D}=D[1-1/(2S)]$ is a renormalized single-ion constant.
Such a renormalization of the anisotropy constant takes us one step beyond the usual harmonic approximation by representing
the distance between the two low energy single-ion levels as 
\begin{equation}
\Delta E = -D[(S-1)^2 -S^2]=D(2S-1)
\end{equation}
in place of  the linear spin-wave result $\Delta E\approx 2DS$. 
Technically, the renormalization factor for $D$ arises from a normal ordering of creation and annihilation operators in 
the quartic boson term originated from $(S^z_i)^2$,
see also Sec.~\ref{MBSWT}.

The energy gap for magnons with zero momenta  is given by:
\begin{equation}
\Delta_T=2S\sqrt{\overbar{D}(4J+\overbar{D})}.
  \label{Delta_T}
\end{equation}
The above expression remains valid  for arbitrary strength of magnetic anisotropy $D/J$
including the standard asymptotic
form  $\Delta_T\approx 4S\sqrt{\overbar{D}J}$ for $D/J\ll 1$ as well as $\Delta_T\approx 2S\overbar{D}$ for 
$D/J\gg 1$.

\subsubsection{Ferromagnet}
\label{Subs_spinS-fm}

The fully polarized ferromagnetic  state  $\ket{\rm FM}$ is an eigenstate of both  $\hat{\cal H}$ and the total
spin operator $S^z$. The corresponding eigenvalues are given by simple classical expressions:
 $E_0 = -(2|J|+D)NS^2$ for the energy and $S^z_\text{tot}=NS$ for the total spin. Here $N=N_L^2$ is the total number of sites and $N_L$ is the linear extension of  a square lattice.  Consequently, $S^z$ remains a good quantum number for the excitations as well.
 
Conventional magnons belong to the subsector  of $S^z_\text{tot}=NS-1$ and therefore each has $S^z=-1$. 
One can still use  harmonic spin-wave theory to compute their dispersion. Alternatively, one can circumvent the Holstein-Primakoff transformation by writing explicitly the one-magnon wave-function  as
\begin{equation}
\ket{1_{\bf k}} = \sum_i e^{-i{\bf k}{\bf r}_i} S_i^-\ket{\rm FM}
\end{equation}
and computing $\hat{\cal H}\ket{1_{\bf k}}$; see, for example, Ref.~\cite{Lifsh-Pit}. 
In either case, the magnon energy is given by
\begin{equation}
\label{et_fm}
 \varepsilon^{T}_\mk{k}=2\overbar{D}S + 4|J|S(1-\gamma_\mk{k})\,.
\end{equation}

\subsection{Longitudinal magnons} 
\label{SLM}

\begin{table*}[tb]
    \begin{tabular}{c|c|c|c|c}
    \hline
    \multirow{3}{*}{~~$S$~~} &  \multirow{3}{*}{$J_\parallel$} & \multirow{3}{*}{$J_\perp$} & 
            \multicolumn{2}{c}{\multirow{2}{*}{$L$-magnon energy }} \\ 
     & & & \multicolumn{2}{c}{ } 
     \rule{0pt}{-10ex} \rule[-1.5ex]{0pt}{0pt}\\ 
     \cline{4-5}
     & & & Ferromagnet & Antiferromagnet \\ 
     \hline\hline
$1$ \rule{0pt}{5ex} \rule[-3ex]{0pt}{0pt} &  
$\displaystyle 4J + \frac{J^2}{D}  -  \frac{1}{2}\frac{J^3}{D^2}$ & 
$\displaystyle-\frac{J^2}{D} + \frac{1}{2}\frac{J^3}{D^2}$ &   &   \\
\cline{1-3}
$\displaystyle\frac{3}{2}$ \rule{0pt}{5ex} \rule[-3ex]{0pt}{0pt} & 
$\displaystyle 9J+\frac{9}{8}\frac{J^2}{D}-\frac{9}{32}\frac{J^3}{D^2}$ & 
$\displaystyle \frac{9}{16}\frac{J^3}{D^2}+\frac{45}{64}\frac{J^4}{D^3}$  &
~~$\varepsilon^L_{\bf k} = 2(|J_\parallel|+ J_\perp\gamma_\mk{k}\bigr) $~~  &
~~$\varepsilon^L_{\bf k} = 2\sqrt{J^2_\parallel-J^2_\perp\gamma_\mk{k}^2}$~~ \\  
\cline{1-3}
$2$ \rule{0pt}{5ex} \rule[-3ex]{0pt}{0pt} & 
~~$\displaystyle16J+\frac{4}{3}\frac{J^2}{D}-\frac{2}{3}\frac{J^3}{D^2}$~~ & 
~~$\displaystyle -\frac{1}{4}\frac{J^4}{D^3}+\frac{3}{8}\frac{J^5}{D^4}$~~ & 
& \\
\cline{1-3}
$\displaystyle\frac{5}{2}$ \rule{0pt}{5ex} \rule[-3ex]{0pt}{0pt}  & 
~~~~~$\displaystyle25J+\frac{25}{16}\frac{J^2}{D}-\frac{25}{32}\frac{J^3}{D^2}$~~~~~ & 
~~~~~$\displaystyle \frac{25}{256}\frac{J^5}{D^4}+\frac{12925}{24576}\frac{J^6}{D^5}$~~~~~ & & \\
\hline     
\end{tabular}
\caption{Exchange parameters in the effective spin Hamiltonian obtained  in powers of $J/D$ for 
different values of $S$. Expressions for the $L$-magnon energy $\varepsilon^L_{\bf k}$ for ferro- 
and antiferromagnetic states are computed in the harmonic spin-wave approximation for the effective 
spin-1/2 model, $\gamma_{\bf k} = \frac{1}{2}(\cos k_x + \cos k_y)$.}
\label{Tab1}
\end{table*}

The strong-coupling expansion in powers of $J/D$  provides a suitable framework
for the description of longitudinal magnons for arbitrary $S\geq 1$. 
Two spin states with $S^z =\pm S$  form  the lowest-energy doublet on every site. These doublets can be associated
 with new pseudo-spin-1/2 variables:
\begin{equation}
\bigl(\ket{+S}, \ket{-S}\bigr)_i \rightarrow  \bigl(\ket{\uparrow},\ket{\downarrow}\bigr)_i\,.
\end{equation}
Spin-spin interactions  in this low-energy subspace are described by an 
effective  Hamiltonian represented in terms of the $s = 1/2$ operators, which are put  
in correspondence with the original spins by
\begin{equation}
s_i^z \sim S_i^z \,,\qquad s_i^\pm  \sim (S_i^\pm)^{2S} \,.
\label{stoS}
\end{equation}
The lowest-order  perturbation processes produce anisotropic  $XXZ$ couplings between pseudospins:
\begin{equation}
\hat{\mathcal{H}}_{\rm eff} = \sum_{\langle ij\rangle} \Bigl[J_\parallel s_i^z s_j^z + J_\perp\bigl( s_i^x s_j^x + s_i^y s_j^y\bigr)\Bigr]
\label{Heff}
\end{equation}
with $J_\parallel \simeq J$ and  $J_\perp/D \simeq (J/D)^{2S}$.
Effective spin-1/2 Hamiltonians similar to (\ref{Heff}) have been previously derived for specific  models and 
 spin values \cite{Damle06,Mardele24,Wierschem12}. Therefore, we skip straightforward intermediate steps and 
present in Table~\ref{Tab1} the coupling parameters $J_\parallel$ and $J_\perp$ for $S=1$, 3/2, 2, and 5/2. 
For each $S$ we include two leading contributions for $J_\perp$ and first three terms for $J_\parallel$. Note that in all 
expressions $J$ can be  positive or negative depending on the initial spin model (\ref{H}). 
In addition to the bilinear terms, 
the higher-order processes beyond those included in derivation of (\ref{Heff}) will  generate multi-spin interactions.
These terms depend on a specific lattice geometry and have additional smallness in $J/D$. Therefore, we exclude
them in the following and base our analysis on Eq.~(\ref{Heff}).

The dominant Ising term establishes the same ferro- or antiferromagnetic structure of pseudospins 
$\langle s_i^z\rangle$ as the magnetic ground state of the original spin-$S$ Hamiltonian (\ref{H}). However, 
the pseudospin excitations described by the effective Hamiltonian (\ref{Heff}) correspond to full flips of 
the original spins rather than to ordinary magnons. These coherently propagating multipolar modes are none 
other than the longitudinal magnons discussed in Sec.~\ref{Intro}. One can use  harmonic spin-wave theory 
for the $s=1/2$  model (\ref{Heff}) to derive their dispersion $\varepsilon^L_{\bf k}$.  The corresponding expressions 
for square-lattice ferro- and antiferromagnets are also included in Table~\ref{Tab1}.

The $L$-magnon energy $\varepsilon^L_{\bf k}$ has  a minimum at ${\bf k}=0$ for both signs of $J$. 
To leading order, the corresponding gap $\Delta_L = \varepsilon^L_0$ is given by  the classical energy of 
a reversed spin $\Delta_L\approx 8|J|S^2$ for both states. In contrast, the $L$-magnon bandwidth differs 
substantially between the two magnetic structures: it is larger for a ferromagnet, $\Delta \varepsilon_L=2 J_\perp$, 
than for an antiferromagnet, $\Delta \varepsilon_L\approx J_\perp^2/J_\parallel$ .

For $J/D\to 0$, the $L$-magnons  are the lowest energy excitations and have  infinite life times. They 
contribute in a standard way to the pseudospin correlation function
\begin{equation}
s^{\alpha\beta}({\bf q},\omega) = \int_{-\infty}^\infty \textrm{d}t\, e^{i\omega t} \langle s^\alpha_{\bf q}(t) s^\beta_{-\bf q}\rangle\,.
\label{sab}
\end{equation}
To be specific, the  transverse spin correlator displays quasiparticle delta peaks
$s^{\perp}({\bf q},\omega) \simeq \delta(\omega-\varepsilon^L_{\bf q})$ for both  ferro- and antiferromagnetic 
states. The longitudinal structure factor for an antiferromagnet  also includes a two-magnon contribution 
\begin{equation}
s^{zz}({\bf q},\omega) =  \pi  \sum_{\bf k}  (u_{\bf k} v_{\bf k-q} + v_{\bf k} u_{\bf k-q})^2 
\delta (\omega - \varepsilon^L_{\bf k} - \varepsilon^L_{\bf k-q}),
\label{szz}
\end{equation}
where $u_{\bf k}$, $v_{\bf k}$ are  the Bogolyubov coefficients \cite{Heilmann81,Canali93,Mourigal10}.
Because of the operator relation (\ref{stoS}), the longitudinal dynamical structure factor of original spins
$S^{zz}({\bf q},\omega)$ features the same $L$-magnon continuum (\ref{szz}). 
The spin selection rules do not restrict  the $L$-magnon response in $S^{zz}({\bf q},\omega)$ for any $S\geq 1$
since the two-particle states in the continuum (\ref{szz}) have $S^z_{\rm tot} =0$.
Spin dependence appears indirectly via $v_{\bf k}\sim J_\perp/J_\parallel \simeq  (J/D)^{2S}$. Hence,
 the continuum intensity (\ref{szz}) decreases with increasing $S$,
 but remains finite as long as the longitudinal magnons form a band of coherent  excitations.

In contrast, the presence of $L$-magnon peaks  in $S^{\perp}({\bf q},\omega)$ is forbidden by the dipole 
selection rules. However,  additional weak spin-orbit terms in real  materials that break the $U(1)$ rotation 
symmetry of (\ref{H}) will mix magnetic excitations with different $S^z$ allowing for {\it weak} $L$-magnon 
peaks  in $S^{\perp}({\bf q},\omega)$. Such a mechanism explains the observation of $L$-magnons  
in ESR and inelastic neutron-scattering experiments on the $S=1$ antiferromagnet FeI$_2$ 
\cite{Fert78,Katsumata00,Legros2021,Bai2021} and in the ESR experiments on 
the $S=2$ antiferromagnets FePS$_3$ and FePSe$_3$  \cite{Wyzula22,Mardele24}.

As $J/D$ increases, the energies of  transverse magnons decrease and they rapidly become the excitations 
of lowest energy. The corresponding level crossing takes place for a value $(|J|/D)_{c1}$ that  can be 
estimated from  $\Delta_L = \Delta_T$. Approximating  the  energy gap for ordinary magnons by 
$\Delta_T\approx 2\overbar{D}S$ and the $L$-magnon gap by $\Delta_L\approx 8|J|S^2$ as seen above, 
we obtain 
\begin{equation}
(|J|/D)_{c1} \simeq 1/(4S) \,.
\label{Jc1}
\end{equation}
The $L$-magnons persist as well-defined excitations for a range of values of $J/D$, before eventually disappearing 
for a  particular value $(|J|/D)_{c2}$. Indeed, for $D=0$ the Heisenberg ferro- and antiferromagnets have no sharp 
excitations with $S^z=2S$ \cite{Manousakis91}. The destruction of $L$-magnons with $S^z=2S$ proceeds as 
a decay into low-energy excitations with smaller $S^z$ values. The corresponding decay processes are not explicitly 
included in the effective Hamiltonian. We treat the $L$-magnon decay in the following Section for the $S=1$ model 
by employing the multiboson representation of spin operators. Still, a threshold value $(|J|/D)_{c2}$, above which 
$L$-magnons acquire a finite life time upon entering the  $2S$-magnon continuum, can be estimated from 
$\Delta_L = 2S\Delta_T$. This  yields
\begin{equation}
(|J|/D)_{c2} \simeq 1/2 
\label{Jc2}
\end{equation}
for the square lattice model (\ref{H}). For more accurate estimates of $(J/D)_{c1}$  and $(J/D)_{c2}$ one 
has to take into account dispersions for both types of magnons. 

We conclude this section by commenting on the sign alternation of  $J_\perp$ for integer and half-integer 
spins of  the original model  with $J>0$, see Table~\ref{Tab1}. This sign alternation is related to the even or 
odd powers of the perturbation expansion, where the full spin-flip processes initially occur for each value of $S$. 
The sign of the transverse coupling plays no role in zero field in the Ising antiferromagnetic state, where
the excitation energy is quadratic in $J_\perp$ (last column, Table~\ref{Tab1}). However, it becomes important in 
the polarized  state above the saturation magnetic field $H_s$, where the sign of $J_\perp$  determines the gap position 
(${\bf k}_0$) in the excitation energy $\varepsilon^L_{\bf k}$  of longitudinal magnons.  At $H=H_s$, the gap 
closes and $L$-magnons with ${\bf k}= {\bf k}_0$ begin to condense. This transition has been recently observed in 
the $S=1$ triangular antiferromagnet Na$_2$BaNi(PO$_4$)$_2$ \cite{Sheng25}. Independently of the lattice structure, 
the multipolar state at $H<H_s$ has ${\bf k}_0 =0$ for integer spins ($J_\perp<0$).  For half-integer spins ($J_\perp>0$), 
the high-field  multipolar state is instead modulated with a finite ${\bf k}_0$, which is equal to  $(\pi,\pi)$ for the square-lattice 
model considered here. The ordering wavevector variation with $S$  is an interesting example of the spin-parity effect 
in two-dimensional quantum magnetism.

\section{$S=1$ antiferromagnet}
\label{AFM}

\begin{figure}[tb]
{\centering
\includegraphics[width=0.9\linewidth]{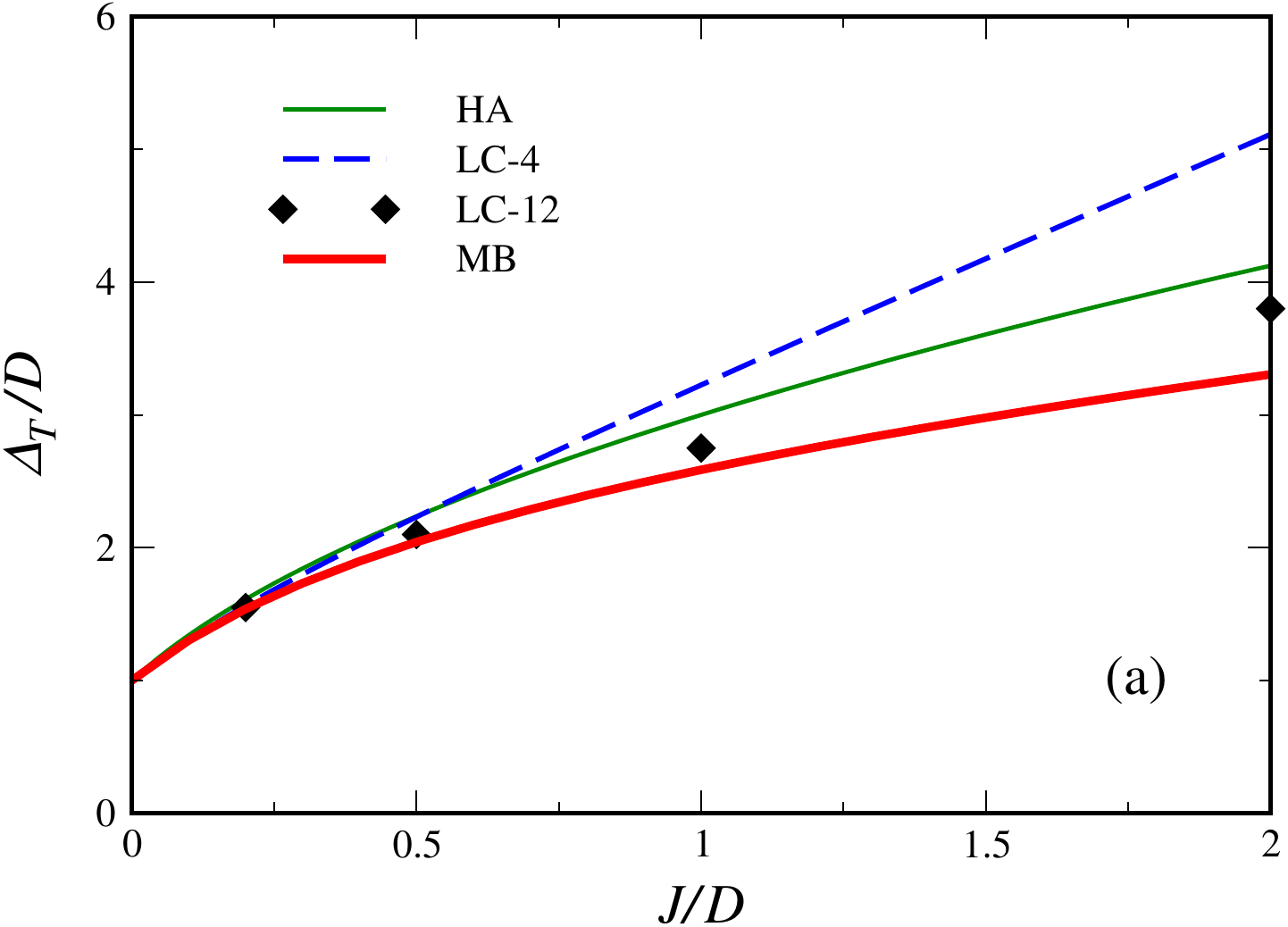}
\vskip 4mm
\includegraphics[width=0.9\linewidth]{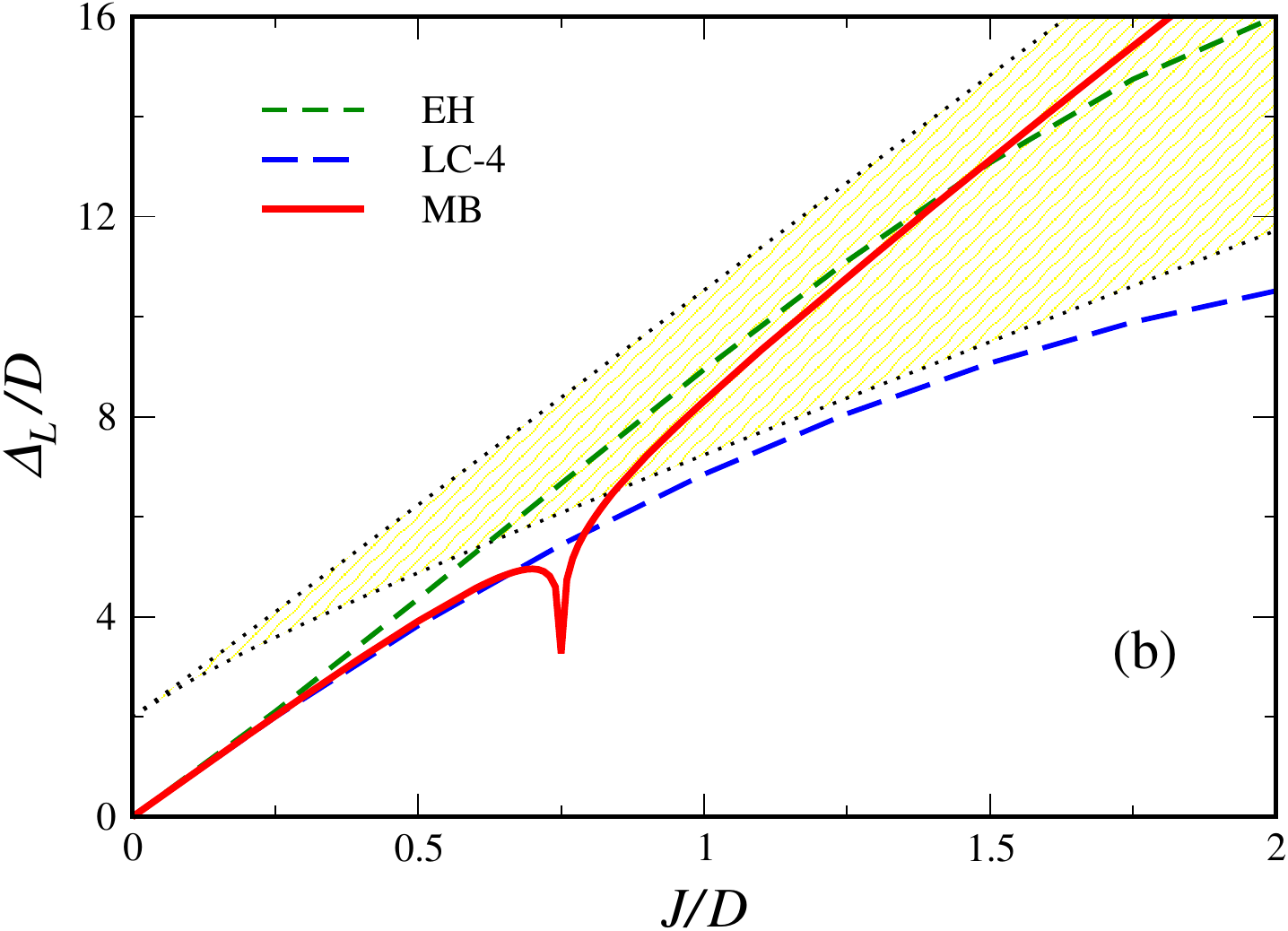}
}
\caption{Energy gaps for $T$-magnons (a) and $L$-magnons (b) for the $S=1$ 
square-lattice antiferromagnet as a function of $J/D$. Legend abbreviations  
 are as follows: HA denotes the harmonic approximation,
EH  is for the effective-Hamiltonian approach, LC-4 is for the fourth-order linked-cluster data, 
LC-12 is for the twelfth-order linked-cluster results of Ref.~\cite{Oitmaa08}, and MB is for the multiboson 
theory including nonlinear quantum corrections. The lightly-shaded region in (b) indicates 
the two-magnon continuum.}
    \label{Gaps}
\end{figure}

Here we discuss two alternative theoretical approaches that complement the effective Hamiltonian method as 
presented in Sec.~\ref{SLM}. The first  is the linked-cluster expansion (Sec.~\ref{subsec-linked-cl}), and 
the second is the multiboson spin-wave theory (Sec.~\ref{MBSWT}).  In principle, both methods can be used 
for an arbitrary spin $S$. However, the analytical calculations are most tractable for small spin values. We shall 
therefore set $S=1$ from now on. Furthermore, in this section we  treat only  the antiferromagnetic model 
(\ref{H}),  with $J > 0$.  The results for the excitation spectra obtained with the use of different methods 
are illustrated in Figs.~\ref{Gaps} and \ref{fig_afm_LM_J-D1}, and further discussed  in Sec.~\ref{subsec-afm-comp}.

\subsection{Linked-cluster expansion}
\label{subsec-linked-cl}

The linked-cluster expansion for quantum spin systems at $T=0$ is a variant of  real-space perturbation theory   \cite{Marland81,Gelfand96,Gelfand00,oitmaa2006book}. In this approach, the energy corrections to  order $n$, 
for both the ground state and the low-lying excited state, are obtained as an extensive sum of the contributions 
from all connected subgraphs (linked clusters) of size up to $n$ bonds \cite{oitmaa2006book}. Selected properties 
of the spin-1 square-lattice antiferromagnet with the easy-axis single-ion anisotropy have been previously computed
to the twelfth order of the linked cluster expansion in $J/D$ \cite{Oitmaa08}. 
Here we obtain  analytical expressions for the ground-state and the $T$- and $L$-magnon excitation energies 
to fourth order, which can be used for comparison with other approaches. 

To construct  the $J/D$ expansion we split the spin Hamiltonian (\ref{H}) into an unperturbed part
\begin{equation}
    \hat{H}_0=J\sum_{\langle ij\rangle}S_i^zS_j^z-D\sum_i ({S}_i^z)^2
    \label{LC_H0}
\end{equation}
diagonal in $S_i^z$, and the transverse coupling, considered as a perturbation 
\begin{equation}
    \hat{V}=\frac{J}{2}\sum_{\langle ij\rangle}(S_i^+S_j^-+S_i^-S_j^+).
    \label{LC_V}
\end{equation}
The ground state is the Néel state with the unperturbed energy per spin:
\begin{equation}
 E_0=-2J-D.   
\end{equation}

The $L$-magnon in this real-space framework is a single-particle excitation formed by a reversed spin 
$\ket{\pm1 }\rightarrow\ket{\mp1}$. The first hopping process for this excitation is of fourth order, involving 
second-nearest neighbors. Accordingly, the leading order of the $L$-magnon dispersion is obtained by considering
processes in clusters with up to four connected bonds. These are dimers, trimers, open tetramers, and squares. 

The preliminary step in this approach is to compute the ground-state energy $E_{\rm g.s.}$ to the same order, 
using the same clusters. Defining $j=J/D$, we obtain
\begin{eqnarray}
\frac{E_{\rm g.s.}-E_0}{D} & = & - \frac{2j^2}{2+7j}-\frac{j^4}{(2+7j)^2}\,
\biggl(\frac{1}{6j} + \frac{4}{1+3j}
\nonumber \\ 
& & + \frac{12}{1+6j} - \frac{46}{2+7j} +\frac{56}{4+13j}\biggr)\,.
\label{Egs_linked-cl}
\end{eqnarray} 
This result compares well to the twelfth-order expansion of Ref.~\cite{Oitmaa08}, showing no more than 1.2\%\ error for $j=1$. 
 
Above the ground state, we obtain the $L$-magnon energy $\varepsilon^L_{\bf k}$ from the sum of all transition amplitudes involving a reversed spin that propagates to its next-nearest neighbors. These amplitudes are read out straightforwardly from the effective Hamiltonian associated to each cluster, see Ref. \cite{oitmaa2006book}. Subsequently, we have
\begin{eqnarray}
\label{el_lc_afm}
\frac{\varepsilon^L_{\bf k}-\varepsilon_0}{D} & = &   - \frac{2j^3}{(2+5j)^2}\Bigl[
(1+4j)\gamma_{\bf k}^2 + 
\\
&  & \mbox{\qquad\qquad\qquad} + 4j\,\frac{3+4j}{1+4j}\cos{k_x}\cos{k_y}\Bigr],
\nonumber
\end{eqnarray}
where
\begin{eqnarray}
&& \frac{\varepsilon_0}{4D} = 2j+ \frac{4j^2}{2+7j} - \frac{3j^2}{2+5j} 
-\frac{j^4}{(2+5j)^2}\,\biggl(\frac{9}{8j} + \frac{11}{2}
\nonumber \\
&&\mbox{\ \ } + \frac{2}{1+4j} - \frac{3}{2+5j} + \frac{7}{2+7j} + \frac{7}{2+10j}\biggr)
  \\
&&\mbox{\ \ }+ \frac{j^4}{(2+7j)^2}\,\biggl(\frac{1}{3j} + \frac{4}{1+3j} + \frac{11}{1+6j} 
- \frac{61}{2+7j}  
\nonumber \\
&&\mbox{\ \ }
- \frac{7}{2+10j} + \frac{56}{4+13j} \biggr) -
\frac{7j^4}{(1+5j)(2+5j)(2+7j)}.
\nonumber 
\end{eqnarray}

\noindent We note that the term $\propto{\cos{k_x}\cos{k_y}}$ stems from the effect of a closed loop in a square. We also note that $\varepsilon^L_{\bf k}$ is minimal at $k=0$. 

The calculation of the $T$-magnon energy to fourth order follows from similar arguments. Here we consider a single spin flip, $\ket{\pm1}\rightarrow\ket{0}$, hopping on dimers and trimers. The resulting energy is
\begin{equation}
\label{et_lc_afm}
\frac{\varepsilon^T_{\bf k}-\tilde{\varepsilon}_0}{D} =  - \frac{8j^2}{1+3j}\Biggl[ 1 + \frac{j}{60(1+3j)} \Big(  j  + \frac{4-33j}{3(1+3j)}\Big) \Bigg]\gamma_{\bf k}^2, 
\end{equation}
where
\begin{eqnarray}
&& \frac{\tilde{\varepsilon}_0}{D} = 1 + \frac{911j}{270} + j^2\biggl(\frac{16}{2+7j} - \frac{181}{45(1+3j)} \biggl) 
\nonumber \\
&&\mbox{\ \ } + \frac{4j^3}{3} \bigg( \frac{1}{(2+7j)^2} - \frac{1}{20(1+3j)^2} \bigg) + \frac{j^4}{2(1+3j)^3} \nonumber  \\
&&\mbox{\ \ } + \frac{j^4}{(2+7j)^2}\,\biggl(\frac{9}{1+6j} - \frac{13}{2+7j} \biggr).
\end{eqnarray}

With these results we may refine the estimates (\ref{Jc1}) and (\ref{Jc2}). The $L$-magnons here are the lowest one-particle excitation for $J/D\le(J/D)_{c1}\simeq 0.18$. They enter the onset of the two-particle continuum for $(J/D)_{c2} \simeq 0.62$. Further analyses of the results are given in Sec.~\ref{subsec-afm-comp}.

\subsection{Multiboson spin-wave theory}
\label{MBSWT}

Several  versions of the multiboson representation of spin operators have been used in the past for different 
problems in quantum magnetism. These include
the spin-dimer singlets \cite{Harris73,Sachdev90,Sommer01},  magnets with the dominant 
single-ion anisotropy \cite{Papa90,Kohama11,Romhanyi12}, and the $SU(N)$ spin models
\cite{Chubukov90,Toth10}.
The basic idea of all these approaches comes from  Hubbard's work
on narrow electron bands in solids  \cite{Hubbard65}.  The objective is to construct the solution of a lattice problem
from the basis states, which take into account strong local correlations. 
In our case, the eigenstates of the single-ion anisotropy term are states with different 
$S^z=\pm 1,0$. Accordingly, we assign 
$\ket{+1}$, $\ket{0}$, and  $\ket{-1}$ to a single-occupant states for three species of bosons $a$, $b$ and $c$, respectively. 
The single-ion anisotropy is then expressed as
\begin{equation}
\label{H_D-mb}
\hat{\cal H}_D = - D \sum_i \bigl(a^\dagger_i a_i + c^\dagger_i c_i\bigr) \,.
\end{equation}
The boson representation of spin operators is constructed so as to reproduce  the matrix
elements between the eigenstates of $\hat{\cal H}_D$:
\begin{equation}
S^z_i  =  a^\dagger_i a_i - c^\dagger_i c_i \,, \quad   
S^+_i  = \sqrt{2}({a}^\dagger_i b_i + b^\dagger_i c_i)  \,,
\label{spin-rep}
\end{equation}
and $S^-_i = (S^+_i)^\dagger$. 
To match the Hilbert space of the physical model,  the boson operators must satisfy the local constraint
\begin{equation}
a^\dagger_i a_i + b^\dagger_i b_i + c^\dagger_i c_i = 1 \,.
\label{constr}
\end{equation}
The above representation of $S=1$ operators is more complicated than the conventional Holstein-Primakoff representation. Nevertheless,  the multiboson representation correctly reproduces the energy spacings between the crystal-field levels on the same site even within the harmonic approximation (see below).

\subsubsection{Boson Hamiltonian}

The representation (\ref{spin-rep}) is used to express the exchange interactions  via the boson operators.
Presence of the antiferromagnetic order is taken into account by applying  the multiboson representation
in the rotated (staggered) local frame with the propagation wavevector ${\bf Q} = (\pi,\pi)$.
Following Ref.~\cite{Sachdev90}, the constraint (\ref{constr}) is enforced via  Lagrange multipliers, by 
adding to the boson Hamiltonian
$\hat{\cal H}_b$ an extra term
\begin{equation}
\hat{\cal H}_L = \sum_i \mu_i\big(a^\dagger_i a_i + b^\dagger_i b_i + c^\dagger_i c_i
-1\big)\,.
\label{mb-constr}
\end{equation}
and imposing $\bra\psi \partial\hat{\cal H}_b/\partial\mu_i\ket\psi=0$ for any boson state $\ket\psi$.
Furthermore, the  translational symmetry allows for a substitution $\mu_i \rightarrow \mu$,
which replaces a set of $N$ constraints  with a single global one. In the rotated local frame, 
the antiferromagnetic ground state corresponds  to a uniform condensate of $a_i$ bosons:
\begin{equation}
\label{a-s-bar}
\braket{a_i}=\braket{a^\dagger_i} = \bar{s}\,.
\end{equation}
The variational parameter $\bar{s}$ is chosen to minimize the ground-state energy. Together with the constraint
on the boson occupation numbers, this yields a set of saddle-point equations on $\mu$ and $\bar{s}$:
\begin{equation}
\frac{\partial E_{\rm g.s.}}{\partial\mu} = 0 \,,\quad \frac{\partial E_{\rm g.s.}}{\partial\bar{s}} = 0 \,.
\label{saddle}
\end{equation}
Note that there is also an alternative way to deal with the constraint (\ref{constr}) by resolving it as
$a_i =  (1-b^\dagger_i b_i - c^\dagger_i c_i)^{1/2}$ and further expanding the square roots \cite{Romhanyi12,Chubukov90}.

The non-condensed $b_i$ and $c_i$ bosons describe transverse and longitudinal magnons with 
$S^z=\pm 1$ and $S^z=\pm 2$, respectively. Regrouping terms according to the number of  $b_i$ and $c_i$
operators, we obtain the boson Hamiltonian as
\begin{equation}
\hat{\mathcal{H}}_b = \hat{\mathcal{H}}_0 + \hat{\mathcal{H}}_2 + \hat{\mathcal{H}}_3 + \hat{\mathcal{H}}_4 \,.
\label{Ham0+2+3+4}
\end{equation}
Here, the constant term is
\begin{equation}
\hat{\mathcal{H}}_0= - N\bigl[2J\bar{s}^4 - \mu(\bar{s}^2-1) + D\bar{s}^2\bigr]
\label{Hmb0}
\end{equation}
The quadratic, cubic, and quartic terms  are  given, respectively, by
\begin{eqnarray}
\hat{\cal H}_2 & = &
\sum_{i} \Bigl[ \mu b_i^\dagger b_i\! +\! (4J\bar{s}^2\! +\!\mu\!-\!D)\,c_i^\dagger c_i\Bigr] 
\nonumber \\
&  & \mbox{\qquad} 
- J \bar{s}^2 \sum_{\langle ij\rangle}{(b_i b_j + b_i^\dagger b_j^\dagger)} \,, 
\nonumber \\ 
\hat{\cal H}_3 &= & -2J\bar{s}\sum_{\langle ij\rangle}{b_i^\dagger (c_j^\dagger + c_i) b_j}\,, 
\label{Hmb}  \\
\hat{\cal H}_4 &= &-J\sum_{\langle ij\rangle}\bigl[\bigl(b_i^\dagger b_j^\dagger c_i c_j +
 c_i^\dagger c_j^\dagger b_i b_j  
\bigr) + c_i^\dagger c_j^\dagger c_i c_j \bigr].
\nonumber  
\end{eqnarray}
A lack of small parameter  for Eq.~(\ref{Hmb})
and similar multiboson Hamiltonians \cite{Harris73,Sachdev90,Sommer01,Papa90,Kohama11,Romhanyi12}
poses a problem for any perturbative treatment  of magnon interactions described by
$\hat{\mathcal{H}}_3$ and $\hat{\mathcal{H}}_4$. In our case, the quantum ground state 
exhibits a classical N\'eel order.  Therefore, we follow a similar sequence of successive approximations
to that in a standard spin-wave expansion. An analog  of  the $1/S$ small parameter
for our problem is  $1-\bar{s}^2$, which measures the strength of zero-point  motion 
in the ground state.

\subsubsection{Harmonic approximation}
\label{subsub-mb-harm}

We start with the classical energy  $\hat{\mathcal{H}}_0$ and obtain, with the help
of (\ref{saddle}), the classical values
\begin{equation}
 \mu_0=D+4J\,,     \qquad   \bar{s}_0=1\,.
\label{mu0}
\end{equation}
The quadratic Hamiltonian ${\hat{\mathcal{H}}}_2$ can be now straightforwardly diagonalized by using 
 Fourier transformation followed by the
canonical Bogolyubov transformation $b_{\bf k} = u_{\bf k} \tilde{b}_{\bf k} + v_{\bf k} \tilde{b}^\dagger_{-\bf k}$, see Appendix.
The ordinary  transverse magnons  are described by the $\tilde{b}_{\bf k}$ boson branch and their excitation energy is
\begin{equation}
\varepsilon^{b}_\mk{k}  =  \sqrt{{\mu}^2-{(4J\bar{s}^2 \gamma_\mk{k})^2}} \,.
 \label{et_mb_h2}
\end{equation}
With  the classical $\mu_0$ and $\bar{s}_0$, the above expression coincides with the linear spin-wave
result (\ref{et_swt_afm}) for $S=1$ with the renormalized $\overbar{D} = D/2$.
Thus, the conventional and the multiboson spin wave theories have the same harmonic spectra 
and the same description of  quantum zero-point oscillations in the semiclassical ground state.

The longitudinal  magnons are described by the $c_{\bf k}$ boson branch and have energy
\begin{equation}
\varepsilon^c_\mk{k} =  \mu - D + 4J \bar{s}^2 \,.
    \label{el_mb_h2}
\end{equation}
They remain fully localized in this approximation and their classical energy 
$\varepsilon_0 =   8J$  comes from four broken exchange bonds around
a fully reversed spin. The many-body effects discussed  next  
determine their dispersion for isotropic  spin models. Nonetheless,
the harmonic multiboson theory has been successfully used for a theoretical
description of the excitation spectra in FeI$_2$ \cite{Legros2021}.  Such an approach is possible
 because due to an absence of spin conservation in this anisotropic material, the longitudinal magnons  
 hybridize with the transverse magnons and acquire their dispersion already in the harmonic approximation.

\subsubsection{Quantum corrections}
\label{subsub-renorm-mb}

In the standard spin-wave theory,  quantum corrections to the classical harmonic spectra  are determined by 
two contributions: (i)  the renormalization of  the ground state  and (ii) the anharmonic terms in the boson Hamiltonian.
Such calculations have been performed using the $1/S$ expansion, for example for the Heisenberg square-lattice antiferromagnet in an applied field \cite{MZH99,Mourigal10} and for the Heisenberg helimagnet with strong quantum fluctuations \cite{Chubukov84,Dalidovich06,Du15}.
 
In our case the leading quantum correction to the ground state energy comes from the zero-point oscillations
obtained after diagonalization of $\hat{\cal H}_2$:
\begin{equation}
\Delta E_{\rm g.s.} = \frac{1}{2N}\sum_\mk{k}{\Bigl[\sqrt{{\mu}^2-{(4 J \bar{s}^2 \gamma_\mk{k})^2}}-\mu\Bigr]}.
 \label{Egs_zp}
\end{equation}
Adding this contribution to the classical energy,  we obtain from the saddle point condition
 (\ref{saddle}) the equations for the renormalized  $\bar{s}$ and $\mu$:
 \begin{eqnarray}
1- \bar{s}^2 & = &  \frac{1}{2N}\sum_\mk{k}\Bigl[ \frac{\mu}{  \sqrt{{\mu}^2-{(4J\bar{s}^2 \gamma_\mk{k})^2}} } -1
\Bigr],
\label{self}
\\
\mu - \mu_0 & = & 4J(\bar{s}^2-1) +\frac{1}{2N}\sum_\mk{k}\frac{\bigl(4J\bar{s} \gamma_\mk{k}\bigr)^2}{ 
\sqrt{{\mu}^2-{(4J\bar{s}^2 \gamma_\mk{k})^2}} } \ .
\nonumber 
\end{eqnarray}
The above equations have to be solved either perturbatively in the spirit of the $1/S$ expansion or self-consistently.
We find that a self-consistent harmonic approach (SCHA) is more practical, providing physical solutions over a wider range of $J/D$.
For example, for $J=D$ we obtain $\bar{s} =0.98$ and $\mu/D= 5.26$  ($\mu_0/D= 5$). 

The   new  $\bar{s}$ and $\mu$  are used to compute  modified  $\varepsilon^{b,c}_\mk{k}$
according to  Eqs.~(\ref{et_mb_h2}) and (\ref{el_mb_h2}).  However, these are not yet the final magnon energies, since the  interaction terms ${\hat{\mathcal{H}}}_3$ and ${\hat{\mathcal{H}}}_4$ generate extra  
contributions that are generally of the same order in  the small parameter  $1-\bar{s}^2$. The  
correction to the energy of $b$ magnons comes from the cubic processes and is expressed as
\begin{equation}\label{Ek-corr3-T}
\delta\varepsilon_\mk{k}^{b}\! =\frac{1}{N} \sum_\mk{q} \biggl[ \frac{\bigl|V_{3}^{(2)}({\bf k},{\bf q})\bigr|^2}
{\varepsilon_\mk{k}^b-\varepsilon^b_\mk{q}-\varepsilon^c_{\mk{k}-\mk{q}}} -
\frac{\bigl|V_{3}^{(1)}({\bf k},{\bf q})\bigr|^2}{\varepsilon_\mk{k}^b + \varepsilon^b_\mk{q} + \varepsilon^c_{\mk{k}+\mk{q}}}
\biggr].
\end{equation}
Further details together with explicit expressions for  vertices $V_{3}^{(n)}({\bf k},{\bf q})$ are provided in Appendix.  
The cubic processes also produce  momentum-dependent corrections
to the energy of  longitudinal magnons:
\begin{equation}
{\delta\varepsilon_\mk{k}^{c}}^{\prime} =\frac{1}{2}\frac{1}{N} \sum_\mk{q} \biggl[ \frac{\bigl|V_{3}^{(3)}({\bf k},{\bf q})\bigr|^2}
{\varepsilon_\mk{k}^c-\varepsilon^b_\mk{q}-\varepsilon^b_{\mk{k}-\mk{q}}} - 
\frac{\bigl|V_{3}^{(1)}({\bf k},{\bf q})\bigr|^2}{\varepsilon_\mk{k}^c + \varepsilon^b_\mk{q} + \varepsilon^b_{\mk{k}+\mk{q}}}
\biggr].
\label{Ek-corr3-L}
\end{equation}
Another contribution to the longitudinal magnon dispersion is determined by the quartic terms:
\begin{equation}
\label{Ek-corr4-L}
{\delta\varepsilon_\mk{k}^{c}}^{\prime\prime} =
-\frac{8(J\Delta_b\gamma_\mk{k})^2}{{\varepsilon_\mk{k}^c}} \ ,
\end{equation}
where $\Delta_b$ denotes the anomalous average:
\begin{equation}
\label{delta_b_mf_afm}
\Delta_b = \langle{b_ib_j}\rangle = 
\frac{1}{2N}\sum_\mk{k}\frac{4\bar{s}^2 J \gamma_\mk{k}^2}
{\sqrt{\mu^2 - (4J\bar{s}^2 \gamma_\mk{k})^2}} \ .
\end{equation}
Another anomalous average $\Delta_c=\langle{c_ic_j}\rangle=0$ vanishes in the harmonic approximation. This results in
the absence of a correction to the energy of $b$ magnons similar to (\ref{Ek-corr4-L}) from the mean-field decoupling
of $\hat{\cal H}_4$.

Results for the magnon spectra obtained by numerical evaluation of the analytical expressions (\ref{Ek-corr3-T}),
 (\ref{Ek-corr3-L}),  and  (\ref{delta_b_mf_afm}) are  presented and discussed in the next subsection.
 We use the renormalized magnon energies  to refine general estimates 
for the critical  couplings (\ref{Jc1}) and 
(\ref{Jc2}) for our model. Specifically,  for $J/D \leq (J/D)_{c1}=0.18$ the
$L$-magnons lie below the conventional $T$-magnon branch and represent  the lowest-energy excitations.
For $(J/D)_{c2}\simeq0.75$, the $L$-magnon band starts to overlap with the two-magnon continuum. As a result,
the longitudinal magnons acquire a finite lifetime $\tau_{\bf k}$ determined by two-magnon decay processees.
In the Born approximation, the decay rate  $\Gamma_\mk{k} =  1/\tau_{\bf k}$ is expressed as
\begin{equation}
\label{Gk-afm}
 \Gamma_\mk{k}  =   \frac{\pi}{2N} \sum_\mk{q}   
 \bigl[V_{3}^{(3)}({\bf k},{\bf q})\bigr]^2\,
 \delta\bigl(\varepsilon^c_\mk{k}-{\varepsilon^b_\mk{q}}-{\varepsilon^b_{\mk{k}-\mk{q}}}\bigr)\,.
\end{equation}

The two-particle decays are responsible for  singularities in the magnon spectra for low-dimensional quantum
antiferromagnets \cite{MZH13,MZH06}. Specifically, in the present case the singularity 
develops near the crossing point denoted by  $\mk{k}^*$. Expanding the denominator of the decay contribution given by the first term in (\ref{Ek-corr3-L}) near the bottom of the two magnon continuum corresponding to $b$ magnons with the momenta
 $\mk{q}^*$ and $\mk{k}^*-\mk{q}^*$, we get: 
\begin{equation}
\varepsilon_\mk{k}^c-\varepsilon^b_\mk{q}-\varepsilon^b_{\mk{k}-\mk{q}}\approx -\mk{v}\cdot\Delta \mk{k}-A q ^2\,,
\end{equation}
where $\mk{v}=\nabla \varepsilon^b_{\mk{k}^*-\mk{q}^*}$ and ${\bf q}\to  \mk{q}^* + {\bf q}$.
Therefore, the on-shell magnon self energy can be approximated by
\begin{equation}\label{sigma_log}
 \Sigma_c({\bf k})\approx   \int\frac{d^2 q}{-v\Delta k -A q^2+i0}
    \simeq \ln{\frac{\Lambda}{\Delta k}}-i\pi\Theta({\Delta k}) \,.
\end{equation}
Here $\Lambda$ is a high-energy cut-off and $\Theta(x)$ is the step function.
The logarithmic divergence is very pronounced in ${\delta\varepsilon_\mk{k}^{c}}^{\prime}$
obtained using Eq.~(\ref{Ek-corr3-L}) for $J/D>0.75$. Corresponding results are further 
analyzed below.

\begin{figure}[tb]
\centering
\includegraphics[width=0.85\linewidth]{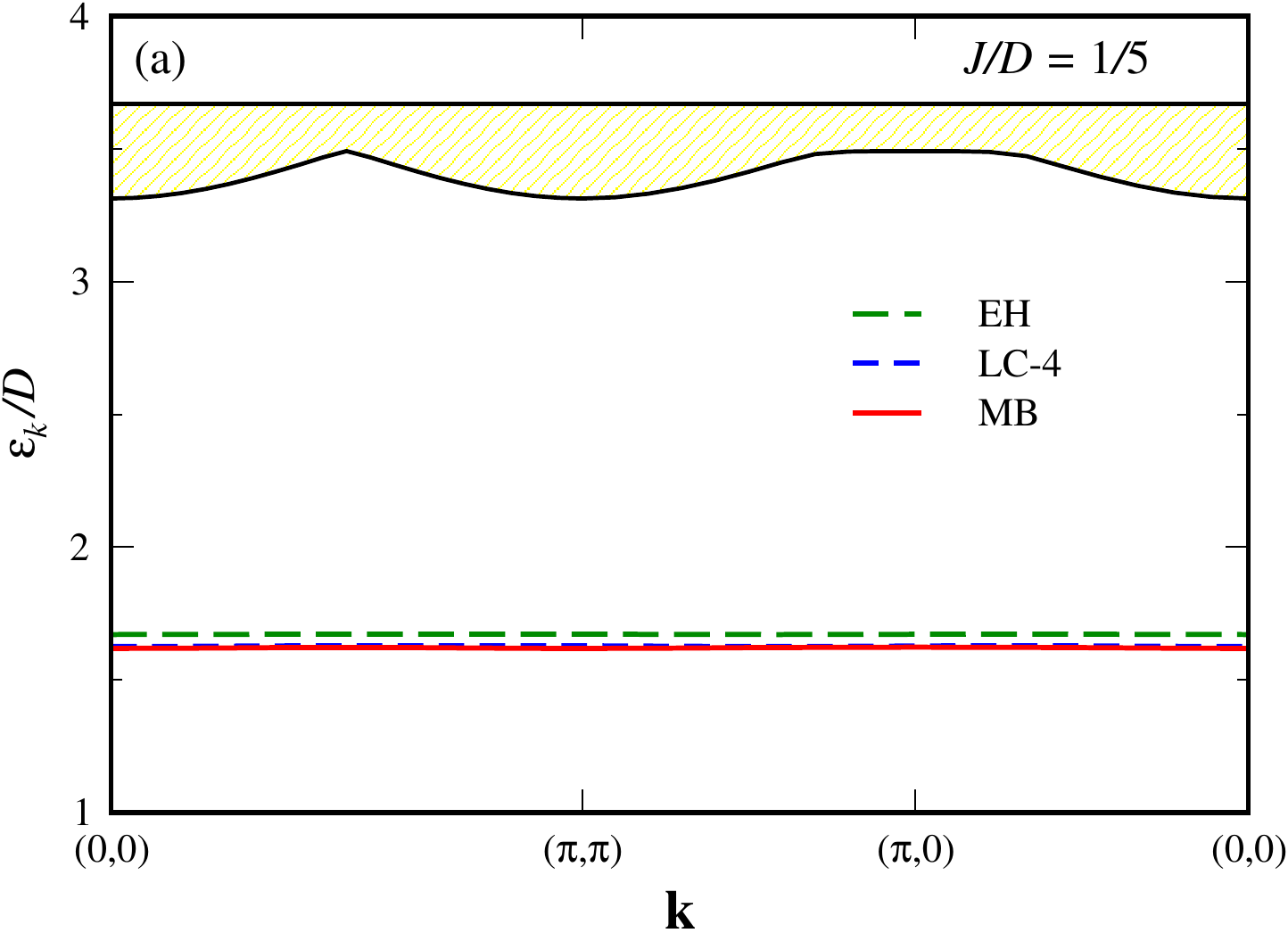}
\label{subf_afm_a}
\vspace{0.75em}
\includegraphics[width=0.85\linewidth]{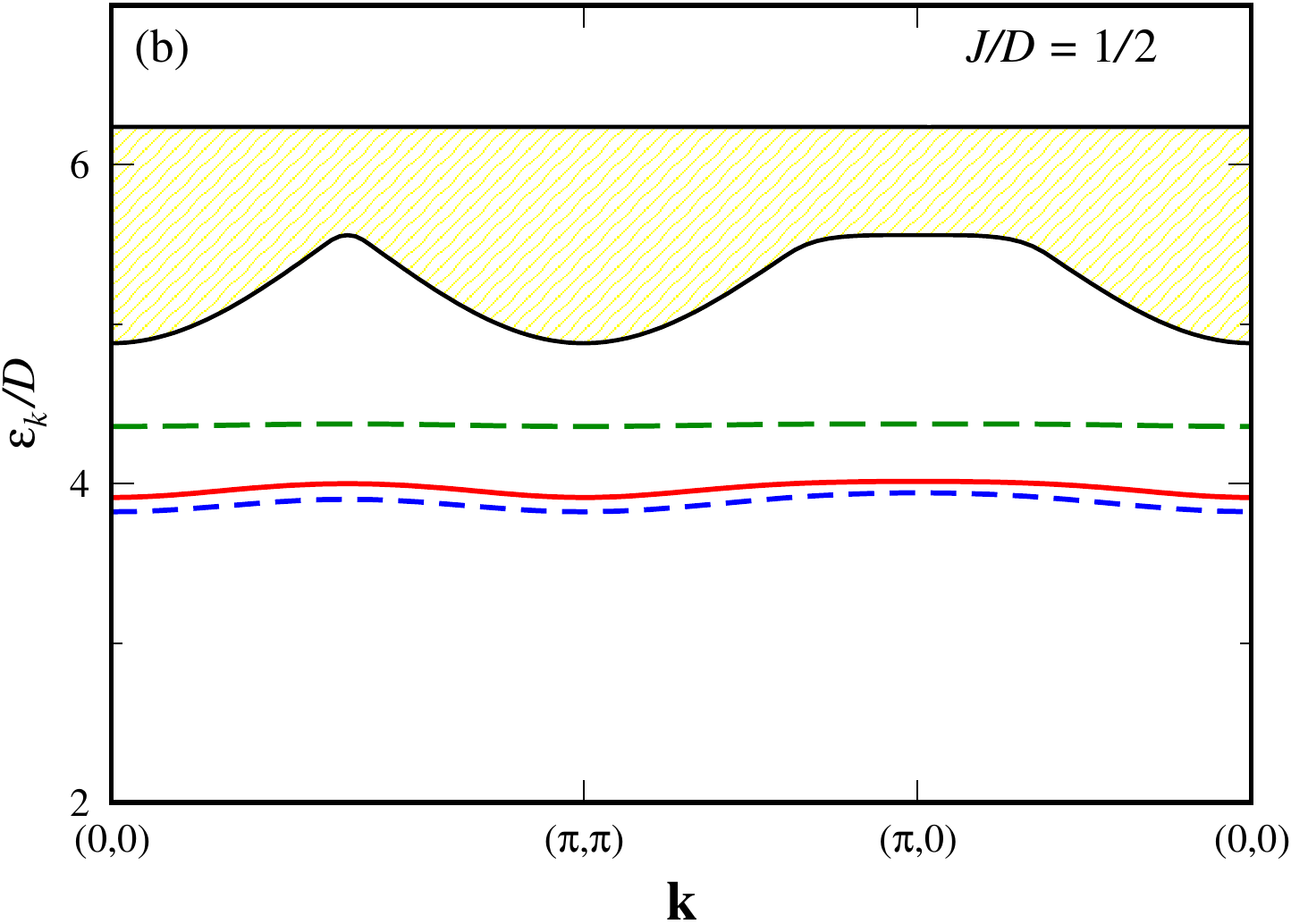}
\label{subf_afm_b}
\vspace{0.5em}
\includegraphics[width=0.85\linewidth]{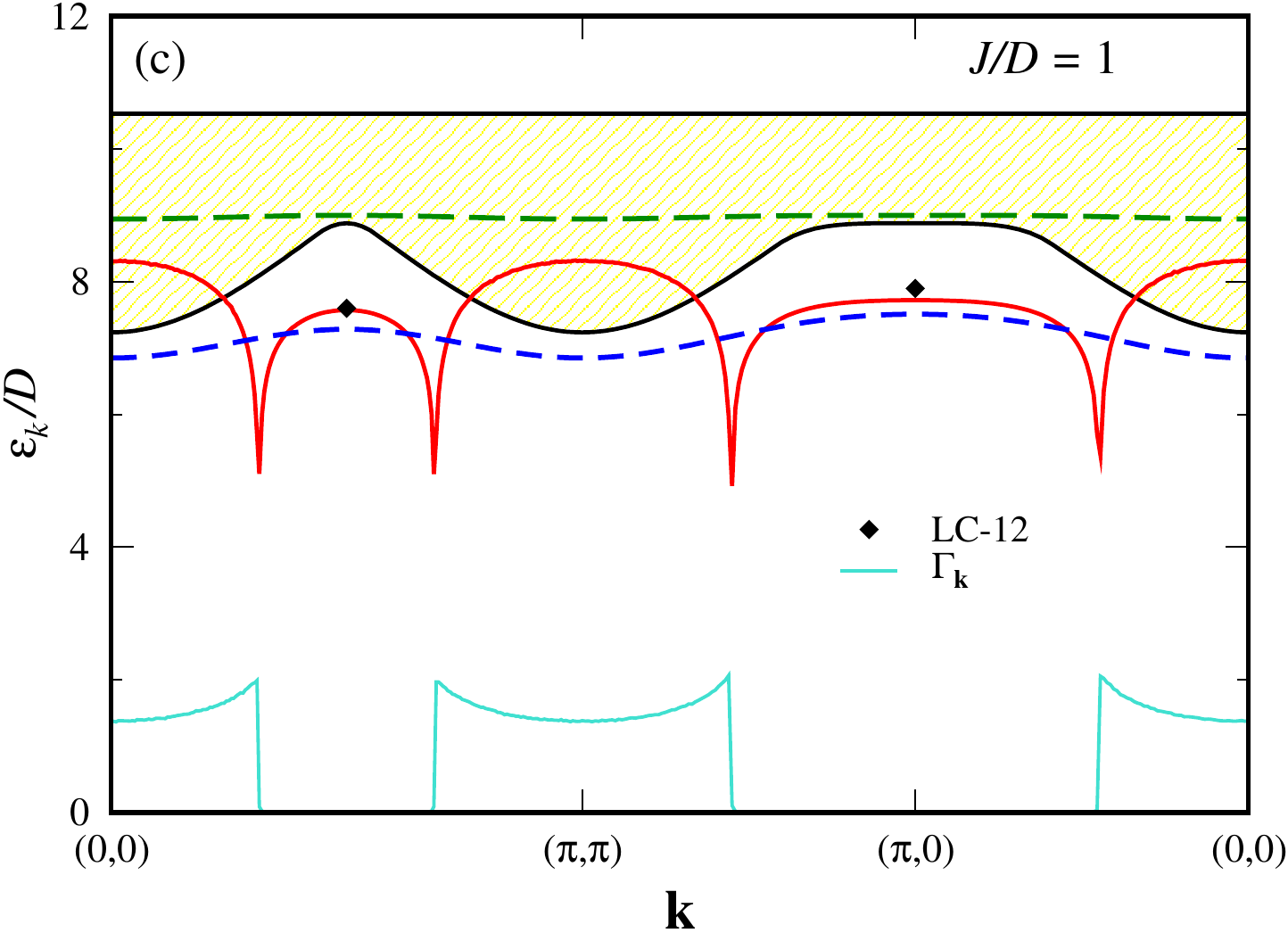}
\label{subf_afm_c}
\caption{
The longitudinal magnon spectra computed by different methods for the antiferromagnet with $J/D=1/5$  (a),  1/2 (b), and 1 (c). Legend abbreviations  are the same as in Fig.~\ref{Gaps}. Lightly shadowed regions denote the two-magnon continuum. Panel (c) includes the $L$-magnon decay rate $\Gamma_\mk{k}$.}
\label{fig_afm_LM_J-D1}
\end{figure}

\subsection{Comparison of results}
\label{subsec-afm-comp}

In the following we discuss the excitation spectra obtained using different theoretical methods. 
The ground-state energy comes out comparatively well for all approaches  except for
the effective Hamiltonian method. That approach loses accuracy at $J/D\agt 0.3$,
 in the parameter range where transverse magnons become the lowest energy excitations 
 and determine the zero-point vacuum fluctuations.

The excitation  gap for ordinary ($T$) magnons $\Delta_T$ is shown  as a function of $J/D$ in Fig.~\ref{Gaps}(a). The twelfth-order linked-cluster expansion results LC-12 computed by Oitmaa and Hamer for  selected values of $J/D$ \cite{Oitmaa08} can be used  as accurate reference values. The
analytical methods agree with the numerical reference values for  small values  of $J/D \alt 0.5$. The fourth-order linked-cluster expansion data, LC-4, start to deviate from
the accurate LC-12 values  for $J/D=1$.  The  harmonic  approximation HA,  which does not depend on
whether  the Holstein-Primakoff or the multiboson representation of spin operators is applied, works better than LC-4 for all  $J/D\alt 1$. The nonlinear quantum corrections  to the harmonic approximation, obtained within the multiboson spin wave theory MB, improve on the agreement with the reference result for  $J/D =1/2$.
In summary, the harmonic theory can give a semi-quantitative description for all $0< J/D < \infty$, whereas
the more elaborate approaches, that take into account nonlinear effects, work reasonably well only for $J/D < 1$.

Figure~\ref{Gaps}(b) presents results for the $L$-magnon gap $\Delta_L$. Because of the lack of accurate 
numerical results, we use our lower-order linked-cluster expansion results LC-4 for comparison with the analytical 
approaches for $J/D \alt 0.75$.  For small values  $J/D \lesssim 0.25$, the three approaches agree with 
one other quantitatively. The effective Hamiltonian results (EH) become less accurate when $J/D$ exceeds $0.3$. The 
harmonic theory yields 
the dispersionless band of longitudinal magnons corresponding to a straight line in Fig.~\ref{Gaps}(b):
 $\Delta_L/D  = 8J/D$. The  MB results include  nonlinear quantum effects and show a better agreement with LC-4 
 theory  up to $J/D\sim 0.5$--0.6. A lightly shaded region in Fig.~\ref{Gaps}(b) corresponds to the extension of the two-magnon continuum for each value of $J/D$. It is constructed using the self-consistent harmonic approximation (SCHA),
 which in our case is only marginally different from the HA results. The crossing point of the unperturbed SCHA 
 branch with the continuum takes place at  $J/D \sim 0.75$. The MB theory displays a sharp dip for the respective value of
 $J/D$, reflecting the  corresponding  logarithmic singularity in the real part of the self-energy (\ref{sigma_log}).

Finally, Fig.~\ref{fig_afm_LM_J-D1} shows the momentum dependence   of the $L$-magnon excitation energy 
for representative values of $J/D$. Overall, the longitudinal magnons have a weak dispersion, which  is, 
nonetheless, noticeable for $J/D\sim 1/2$. The multiboson spin wave theory  is in good quantitative agreement  
with the LC-4 results for small $J/D\alt 0.5$. The effective Hamiltonian approach  gives in this coupling range 
somewhat overestimated energy $\varepsilon^L_{\bf k}$. For larger  $J/D\agt 0.75$, the longitudinal magnon 
branch enters the two-magnon continuum. The multiboson theory is the only analytical approach that can predict 
finite life times for the $L$-magnons in this regime.

Close to the crossing point with the continuum the second-order correction (\ref{Ek-corr3-L})  develops 
a logarithmic singularity as a function of momentum described by Eq.~(\ref{sigma_log}).  Such a behavior is 
an artifact of the Born approximation in conjunction with the low dimensionality of the spin model \cite{MZH13}. 
The quantum correction  $\delta\varepsilon^L_{\bf k}$ is regularized by  higher-order processes. Specifically, 
the singularity from the two-particle  decays is treated by taking into account 
the vertex corrections via a resummation of  most divergent diagrams \cite{Lifsh-Pit,MZH13}.
Analysis of the analytical properties of the magnon self-energy $\Sigma_c(\omega,{\bf k})$ suggests 
that in two dimensions the single-particle branch  $\varepsilon^L_{\bf k}$ enters the continuum tangentially
with  vanishing quasiparticle weight towards the crossing point \cite{MZH06}. 
 
In the bottom panel, Fig.~\ref{fig_afm_LM_J-D1}(c), corresponding to $J/D=1$, we also include two points
corresponding to longitudinal magnons with ${\bf k}=(\pi/2,\pi/2)$ and  ${\bf k}=(\pi,0)$ obtained with the
help of the high-order linked cluster expansion LC-12 \cite{Oitmaa08}. As one can see from the comparison 
with the MB curve, the interacting  multiboson approach remains in agreement with the numerical results 
for relatively large $J/D$ as long as the $L$-magnon branch lies outside of the continuum. The inverse lifetime 
of longitudinal magnons $\Gamma_{\bf k}$ (in units of $D$) is also included to this plot. The ratio 
$\Gamma_{\bf k}/\varepsilon_{\bf k}$ does not exceed $\sim 0.3$ for this value of $J/D$, implying that 
the longitudinal magnons remain as well-defined resonances inside the continuum. The decay rates of 
the longitudinal magnons rapidly grow upon further increase of the coupling constant. Their transformation 
into short-lived excitations throughout the whole Brillouin zone takes place for 
\begin{equation}
(J/D)_{c3}\simeq 1.55 \ .
\end{equation}

\section{$S=1$ ferromagnet}
\label{FM}

We now consider the ferromagnetic model  (\ref{H}) with $J<0$.  Presence of the exact ground state with a
parallel spin alignment along $\pm z$ direction allows for the exact solution of the two-particle problem.
Besides the two-magnon continuum, the corresponding  eigenstates  include  the exchange and the single-ion bound states. For $S=1$, the latter states are identical  to the longitudinal magnons. In the following we obtain the exact 
dispersion of $L$-magnons by solving the two-particle Schr\"odinger equation
and compare it to results of two approximate methods: the effective spin Hamiltonian and the multiboson spin-wave theory.

We denote the fully polarized ferromagnetic state with spins pointing in the $+z$ direction by $\ket{\rm FM}$.
A general state in the two-magnon subsector with $S^z=-2$ can be represented as
\begin{equation}
\ket{2}=\sum_{i,j}{\psi_{ij}\,{S}^-_i{S}^-_j\ket{\rm FM}}
    \label{2magnon}
\end{equation}
with site  summation implied and $\psi_{ij}=\psi_{ji}$. The eigenstate condition  for the state $\ket{2}$ yields    
\begin{equation}
\varepsilon_2 \ket{2}=\sum_{i,j}\psi_{ij}[\hat{\mathcal{H}},{S}^-_i{S}^-_j]\ket{\rm FM} \,,
    \label{E2-commut}
\end{equation} 
where $\varepsilon_2 = E_2 - E_{\rm g.s.}$ is the excitation energy.

Investigation of bound magnon states in ferromagnets based on the Schr\"odinger equation (\ref{E2-commut}) have
 been performed in Refs.~\cite{Wortis63,Hanus63,Torrance1969IBF,Tonegawa71,Mattis86}. 
This approach allows to track a real-space structure of the bound-state wave function as well as its symmetry.
Below, we revisit this method and formulate  Eq.~(\ref{E2-commut}) as  a linear eigenvalue problem for the binding energy. Other approaches to the same problem include  the Bethe ansatz method used for the anisotropic spin chains \cite{Hodgson85,Papa87} and the Green's function  calculations \cite{Silberglitt70}. 

The single-ion bound states ($L$-magnons) are distinguished from the exchange bound states  (EBS) 
by  amplitude ratios: $|\psi_{ii}|\agt|\psi_{i,i+\bm{\delta}}|$ for the former and $|\psi_{ii}|\alt|\psi_{i,i+\bm{\delta}}|$ for the latter, where $\bm{\delta}$ is the  bond vector. In particular, the single-ion bound states always correspond to the eigenstates with the $s$-wave symmetry,
 whereas the exchange bound states  may have various symmetries
 depending on a spin model.

\subsection{Exact solution}
\label{fm-ex}

The operator commutators in Eq.~(\ref{E2-commut}) are computed separately for  $i\neq j$ and for $i=j$.
The obtained linear equations for $\psi_{ij}$  are expressed in  the momentum space with the help of  the 
double Fourier transformation:
\begin{equation}
\label{FT-2}
\psi_{ij} = \frac{1}{N} \sum_{\mk{k}}  e^{i\mk{k}(\mk{r}_i+\mk{r}_j)/2 } \,\psi_{\mk{k}}(\mk{r})\ ,
\end{equation}
with 
\begin{equation}
 \label{FT_psi_r}
\psi_{\mk{k}}(\mk{r}) =  \frac{1}{N}  \sum_{\mk{q}} e^{i\mk{q} \mk{r}}   \psi_{\mk{k},\mk{q}}\  ,
\end{equation}
where $\mk{r}=\mk{r}_j-\mk{r}_i$  is a relative
distance between two spin flips and  $\mk{k}$ is a total momentum of the magnon pair. The end result is
 a system of two equations: 
\begin{eqnarray}
&& \bigl[\varepsilon_2-\varepsilon^{T}_{\mk{k}/2+\mk{q}} - \varepsilon^{T}_{\mk{k}/2-\mk{q}}\bigr]\psi_{\mk{k},\mk{q}}
=  \frac{2J}{N} \sum_\mk{p }  \Bigl[ \cos{q_x}
\label{E2-ij}  \\
&& \mbox{} \times \bigl(\cos{p_x}-\cos{\textstyle \frac{1}{2}k_x}\bigr)
 + \cos{q_y}\bigl(\cos{p_y}-\cos{\textstyle \frac{1}{2}k_y}\bigr)\Bigr] \psi_{\mk{k},\mk{p}} \ ,
\nonumber 
\end{eqnarray}
which describes the intersite hopping $\psi_{ij} \to \psi_{i+\bm{\delta},j}$ and 
\begin{equation}
\frac{1}{N} \sum_{\mk{p}}\Bigl[\varepsilon_2-\varepsilon^{T}_{\mk{k}/2+\mk{p}}-\varepsilon^{T}_{\mk{k}/2-\mk{p}}
+ 2D\Bigr] \psi_{\mk{k},\mk{p}} = 0 \ ,
\label{E2-ii}
\end{equation}
which relates  $\psi_{ii}$ to the nearest-neighbor amplitudes.
Here $\varepsilon^{T}_\mk{k}$ is the one-magnon excitation energy (\ref{et_fm}).
For $D=0$, the second equation  follows directly from the first one.  As a result, one recovers the same eigenstate problem as  for the exchange bound states  in the Heisenberg ferromagnet \cite{Mattis86}. For $D> 0$, the attractive single-ion term in Eq.~(\ref{E2-ii}) creates an `impurity potential'.
This breaks translational invariance for the relative motion of two spin flips and, therefore, precludes
solution of the two-magnon problem by an expansion in a few lattice harmonics.

Below, we solve the eigenstate problem  in the mixed momentum/real-space representation of
the wave function $\psi_{\mk{k}}(\mk{r})$, which is appropriate for bound two-magnon pairs.  
We further use the short-hand notations  $c_{x,y} = 4\cos  k_{x,y}/2$
and define dimensionless parameters
\begin{equation}
d={D}/|J|\,, \quad e_b = \varepsilon_2/J + 8 + 2d - c_x - c_y \,.
\label{prm_d_eb}
\end{equation}
Here $e_b$ denotes the dimensionless binding energy, measured  from the bottom of the two-magnon continuum.
The momentum resummation in Eqs.~(\ref{E2-ij}) and (\ref{E2-ii})  yields
\begin{eqnarray}
\label{E2-psi-Fh}
&& \bigl(e_b + c_x + c_y - c_x\cos{q_x} - c_y\cos{q_y} \bigr)  \psi_{\mk{k},\mk{q}} = 2\cos{q_x} 
 \nonumber \\
&&\times \bigl[ \psi_\mk{k} (\bm{a})\! -\! {\textstyle\frac{1}{4}} c_x\psi_\mk{k}(0)\bigr]\!
+  2 \cos{q_y} \bigl[\psi_\mk{k}(\bm{b})\!- \!{\textstyle\frac{1}{4}} c_y\psi_\mk{k}(0) \bigr] \  \qquad
\label{inter-site} 
\end{eqnarray}
and
\begin{equation}
 \bigl(e_b + c_x + c_y - 2d\bigr) \psi_\mk{k}(0) = 
c_x\psi_\mk{k}(\bm{a}) + c_y \psi_\mk{k}(\bm{b}) \ ,
\label{same-site} 
\end{equation}
respectively,
with $\bm{a}=(1,0)$ and  $\bm{b}=(0,1)$. 

Next, we specify $\mk{r}=(m,n)\ne 0$, with integer $m,n=-N_0,\ldots,N_0$ such that the linear extension
in each direction is $N_L = 2N_0+1$ and
convolute both sides of Eq.~(\ref{inter-site}) with $\cos (\mk{q}\cdot\mk{r})$, summing up over 
 $\bf q$ in the first Brillouin zone. This yields 
\begin{eqnarray}
&&
(e_b+c_x+c_y)\psi_\mk{k}({\bf r}) - \frac{c_x}{2}\bigl[{\psi_\mk{k}({\bf r}+\bm{a}) + 
\psi_\mk{k}({\bf r}-\bm{a})}\bigr]
\nonumber   \\
&& \mbox{} - \frac{c_y}{2}\bigl [{\psi_\mk{k}({\bf r}\!+\!\bm{b})\! + \!\psi_\mk{k}({\bf r}\!-\!\bm{b})} \bigr]
=\delta_{\mk{r},\bm{a}} \bigl [\psi_\mk{k}(\bm{a})\!-\!\frac{c_x}{4}\psi_\mk{k}(0 )\bigr]
\nonumber  \\
&& \mbox{\qquad} +
\delta_{\mk{r},\bm{b}} \bigl[\psi_\mk{k}(\bm{b})-\frac{c_y}{4}\psi_\mk{k}(0)\bigr] \ ,
  \label{gen-psi}
\end{eqnarray}
where $\delta_{{\bf r},{\bf r}'}$ is the Kronecker function and $\psi_\mk{k}(\mk{r})=\psi_\mk{k}(-\mk{r})$
has been used. 

The bound-state solutions  exhibit an exponential decay  of the wave function $\psi_\mk{k}({\bf r})$
with distance  ${\bf r}$. Focusing on these states we can modify the periodic boundary conditions imposed
by the Fourier transformation  and consider Eq.~(\ref{gen-psi})
on finite $N_L\times N_L$ lattices with open boundary conditions. In addition, we rearrange amplitudes
from a two-dimensional plane onto a one-dimensional columnar array of length 
$N=N_L\times N_L$:
\begin{equation}\label{Psi}
    \Psi_\mk{k} = \bigl(\{\psi_\mk{k}(-N_0,n)\},\ldots\{\psi_\mk{k}(0,n)\},\ldots\{\psi_\mk{k}(N_0,n)\}\bigr).
\end{equation}
Equations (\ref{same-site})  and (\ref{gen-psi}) are now put in the matrix form
\begin{equation}
    \!\!\begin{pmatrix}
  \mathbb{P}_0&\mathbb{J}\\
    \mathbb{J}&\mathbb{P}_0&\mathbb{J}\\
    & & ...\\
    &&&\mathbb{J}&\mathbb{P}_1&\mathbb{Q}\\
    &&&&\mathbb{J}&\mathbb{P}_2&\mathbb{J}\\
    &&&&&\mathbb{Q}&\mathbb{P}_1&\mathbb{J}\\
    &&&&&& & ...\\
    &&&&&&&\mathbb{J}&\mathbb{P}_0&\mathbb{J}\\
    &&&&&&&&\mathbb{J}&\mathbb{P}_0
\end{pmatrix}\Psi_\mk{k}=0\ .
\label{R.psi}
\end{equation}
Here, all elements are $N_L\times N_L$ matrices. Matrix $\mathbb{J}$ is proportional to the unit matrix:
$\mathbb{J}=(-c_x/2)\,\mathbb{I}$. Matrix $\mathbb{Q}$ is diagonal with
\begin{eqnarray}
    &&\mathbb{Q}(i,i)=-c_x/2\,,\quad i\ne N_0+1,\nonumber\\
    &&\mathbb{Q}(N_0+1,N_0+1)=-c_x/4\,.
\end{eqnarray}
Tridiagonal matrices $\mathbb{P}_{l=0,1,2}$ have the same subdiagonal elements
\begin{equation}
\mathbb{P}_l(i,i\pm1)=-{c_y}/2\ 
\label{Poff}  
\end{equation}
except for two elements in $\mathbb{P}_2$:
\begin{equation}
    \mathbb{P}_2(N_0+1\pm1,N_0+1)=-{c_y}/4\,.
\end{equation}

The diagonal elements of $\mathbb{P}_0$ are all equal to
\begin{equation}
\mathbb{P}_0(i,i)=e_b+c_x+c_y\ .
\label{P0}  
\end{equation}
The central diagonal element  of $\mathbb{P}_1$ differs  from the above 
\begin{equation}
\mathbb{P}_1(N_0+1,N_0+1)=e_b+c_x+c_y-1\ .
\label{P1}  
\end{equation}
For $\mathbb{P}_2$, three elements are changed with respect to (\ref{P0}):
\begin{eqnarray}
&&  \mathbb{P}_2(N_0+1,N_0+1) = e_b+c_x+c_y-2d \ ,
\nonumber \\
&& \mathbb{P}_2(N_0+1\pm1,N_0+1\pm1)= e_b+c_x+c_y-1\,.
\label{P2}
\end{eqnarray}

The eigenvalue problem for $e_b$ (\ref{R.psi}) can be solved  with the help of  
standard numerical algorithms. Specifically, we have used the Lapack routines 
to compute eigenvalues  outside of the two-magnon continuum for  square clusters with
increasing linear dimension $N_L$. The obtained results exhibit rapid convergence with 
increasing system size, providing four-digit accuracy for any $\bf k$ with $N_L=35$.

For all values of $\mk{k}$  there are at most  three solutions with $e_b>0$, which can be associated with the single-ion
($L$-magnon) and two exchange bound states. Using the Gershgorin circle theorem on eigenvalue bounds for square matrices \cite{Varga}, we obtain: 
\begin{equation}
-2(c_x+c_y)  \le  e_b \le \max\{2d,1\}.
\label{Gersh}
\end{equation}
Since the width of the continuum is  $2(c_x+c_y)$, there are no anti-bound states above the continuum. 
The right-hand side of the previous inequality sets a lower bound for the two-magnon energy:
\begin{equation}
    \varepsilon_2/|J|\ge 8+2d-c_x-c_y-\max\{2d,1\}.
    \label{E2_min}
\end{equation}

\begin{figure}[tb]
\centering
\includegraphics[width=0.9\columnwidth]{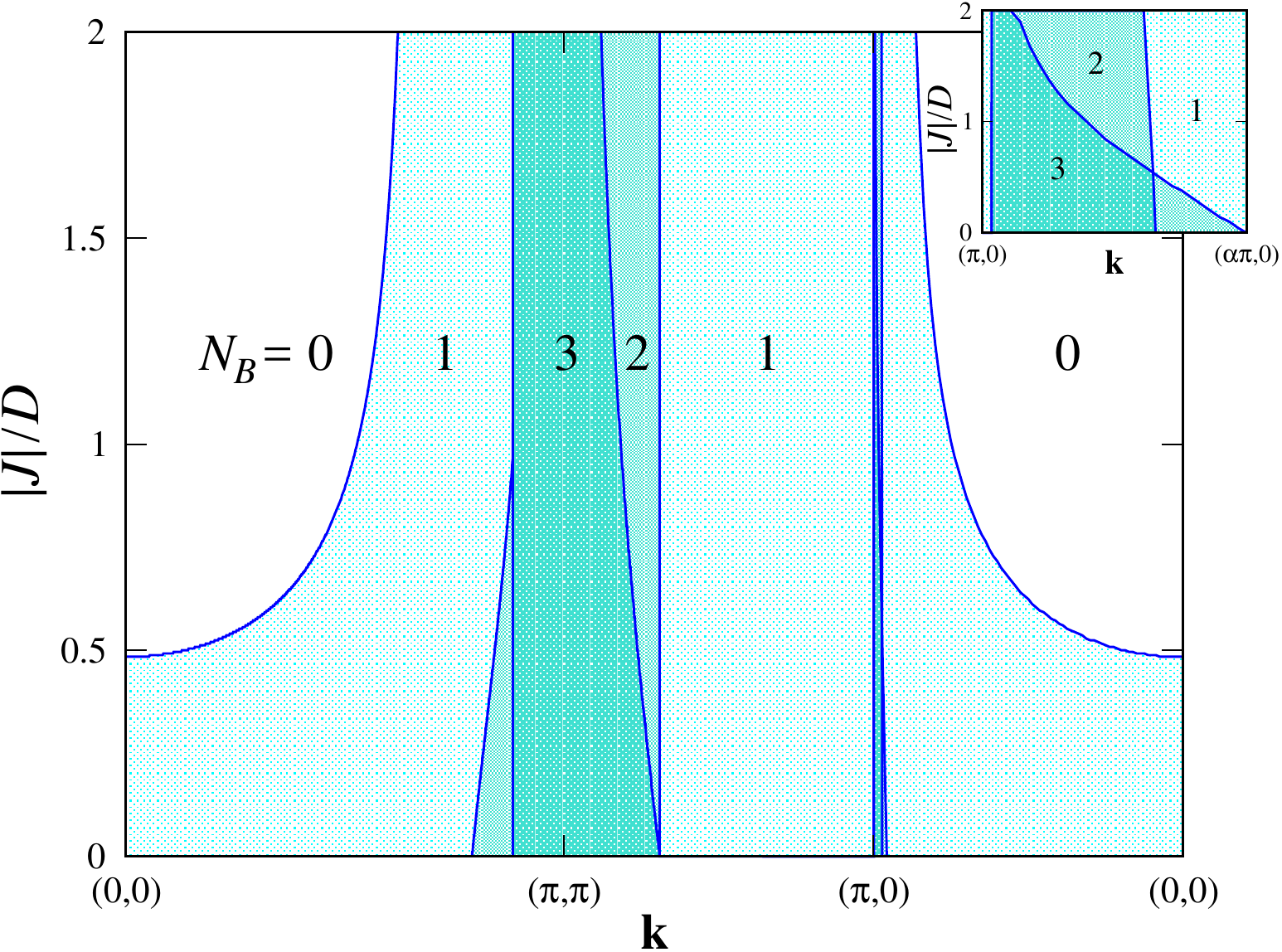}
\caption{Number of bound states for different total momenta $\bf k$ for the easy-axis ferromagnet
on a  square lattice. The horizontal line represents a path in the Brillouin zone. The vertical line
spans a range of $|J|/D$ values. The number of bound states varies for each of  the shaded regions
and is indicated by $N_B$.  The inset zooms in on the region close to $\mk{k}=(\pi,0)$ with $\alpha = 0.986$}. 

\label{Nb}
\end{figure}

The matrix in (\ref{R.psi}) has a  centrosymmetric structure, which leads to special symmetry properties of
its eigenstates. Specifically, the eigenstates  of a centrosymmetric matrix are classified  
into centro-symmetric and skew-symmetric vectors \cite{Andrew73}. The former correspond to the $s$-wave bound-state 
wavefunctions, whereas the latter describe the $d$-wave bound states.

For $k_x=\pi$, the matrix (\ref{R.psi}) 
 acquires a block-diagonal form leading to independent secular equations: 
\begin{equation}\label{det_Pk}
 \det (\mathbb{P}_l) = 0 \ ,\quad l = 0,1,2  \ .
\end{equation}
The Gershgorin theorem applied to  $\mathbb{P}_0$  shows that this matrix has only negative
eigenvalues  corresponding to states in the continuum. Thus, we are left only with two equations for
$l=1,2$. The matrix structure is further simplified for $\mk{k}_0=(\pi,\pi)$ as all matrices $\mathbb{P}_l$ 
become diagonal. In this case, the block  $\mathbb{P}_2$ provides two  solutions with positive $e_b$: 
(i) the single-ion bound state with $e_b=2d$ and  $\psi_{\mk{k}_0}(0)\ne 0$, and 
(ii)  the exchange bound state with $e_b=1$ and  $\psi_{\mk{k}_0}(\pm\bm{b})\ne 0$. 
Matrix $\mathbb{P}_1$ yields a second  exchange bound state with  $e_b=1$ and 
$\psi_{\mk{k}_0}(\pm\bm{a})\ne 0$.  Overall,  there are 
three bound states  with the total momentum $\mk{k}_0$.

The diagram in Fig.~\ref{Nb} demonstrates how the number of bound states  changes upon moving away 
from the $(\pi,\pi)$ point depending on $|J|/D$ value. The single-ion bound states ($L$-magnon branch)
can easily be distinguished from the exchange bound states, as the latter  do not exist for zero total momentum 
of a magnon pair ${\bf k}=0$ for any value of $J$. The obtained results agree with a proposition that an 
anisotropic  ferromagnet in $\mathcal{D}$ dimensions has $\mathcal{D}+1$ bound states \cite{Silberglitt70}.  
Using the developed approach, we can, in fact, prove this theorem  for a $\mathcal{D}$-dimensional hypercubic 
lattice.

\subsection{Comparison of results}

\begin{figure}[tb]
\centering{
\includegraphics[width=0.8\columnwidth]{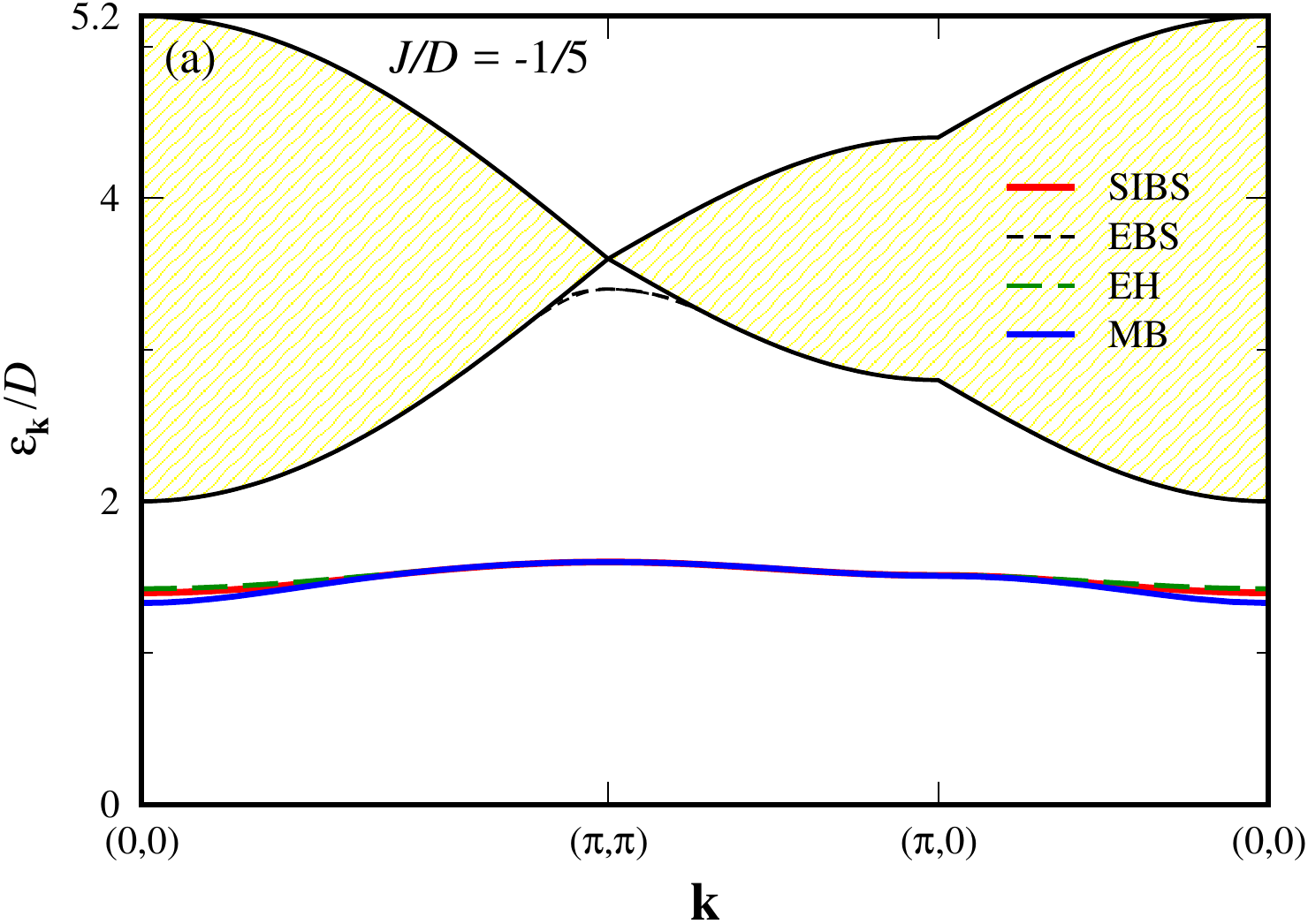}
\vspace{0.5em}
\includegraphics[width=0.805\columnwidth]{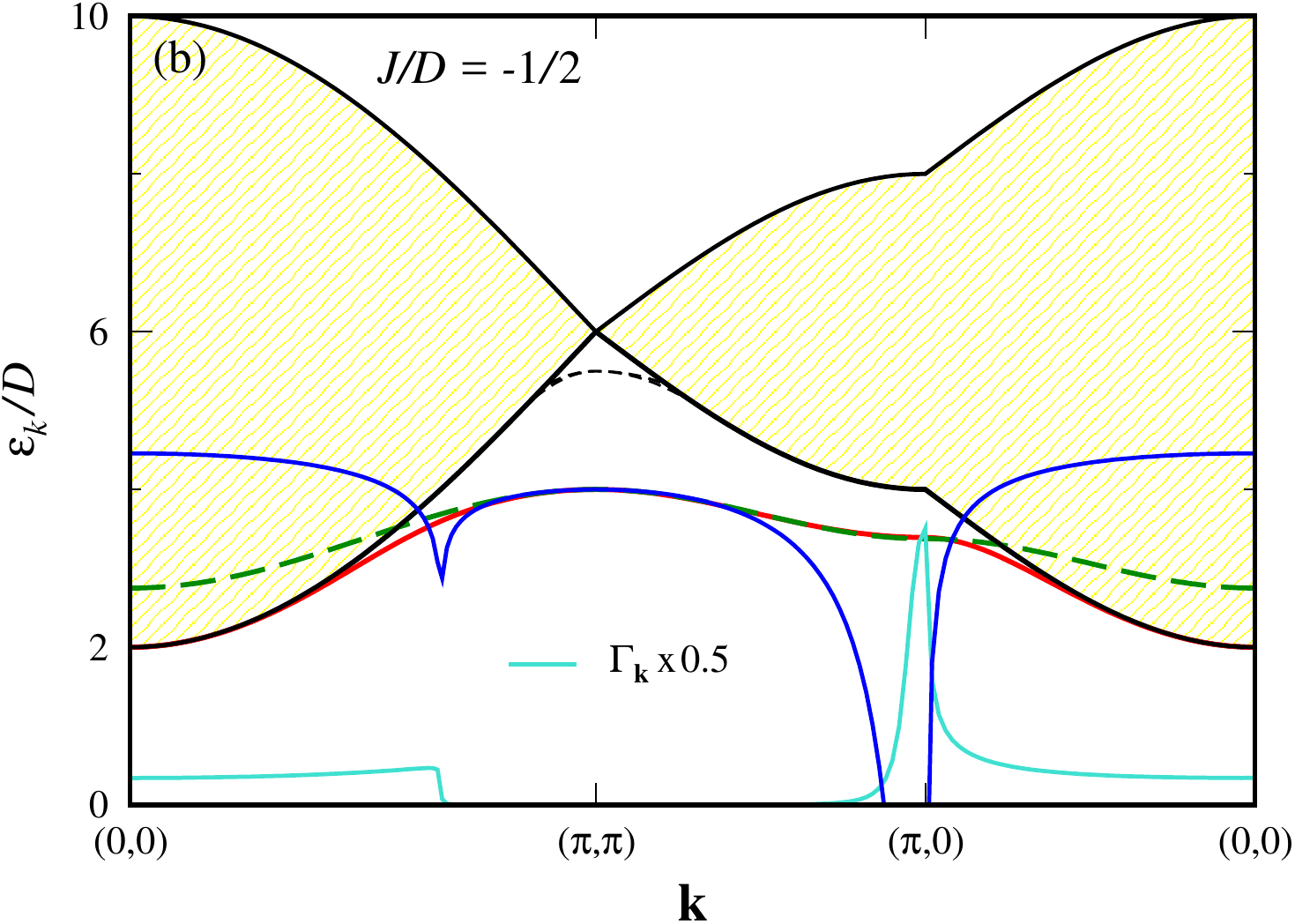}
\vspace{0.5em}
 \includegraphics[width=0.8\columnwidth]{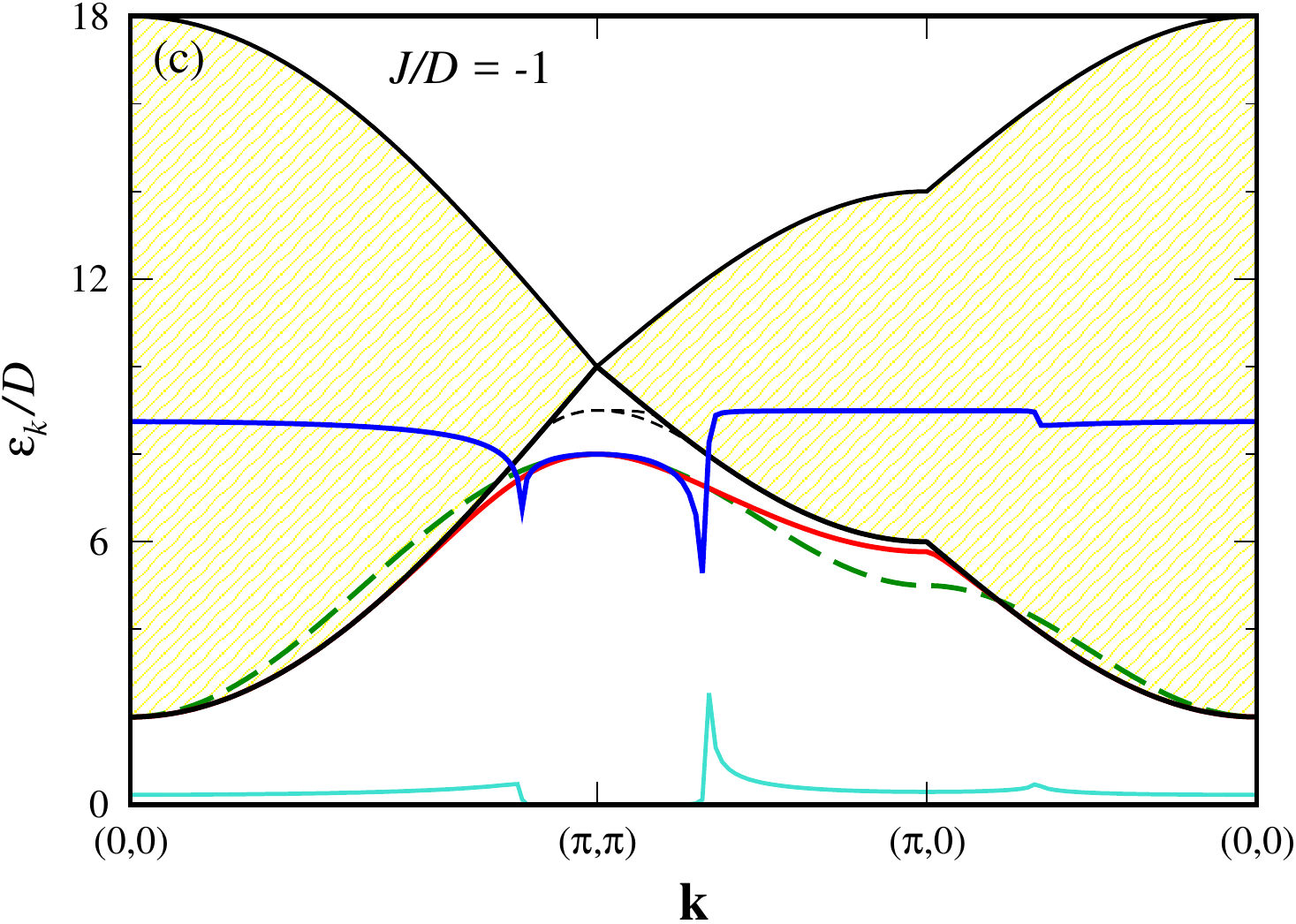}
}
    \caption{
The two-magnon spectrum for the $S=1$ square-lattice ferromagnet. Legend abbreviations are SIBS for the single-ion bound state ($L$-magnon), EBS for the exchange bound state, both from the exact solution, EH for the effective Hamiltonian approach and MB for the multiboson spin-wave theory. 
Panels (b) and (c) include the  $L$-magnon decay rate $\Gamma_\mk{k}$.
}
    \label{fig_fm_LM_J-D1}
\end{figure}

From the exact solution of the two-particle problem, we find that
the $L$-magnon band has always a minimum at ${\bf k}=0$, whereas
the top of the  band is located at  ${\bf k}=\mk{k}_0$ with $\varepsilon_0 = 8|J|$
coinciding  with the classical energy of a fully reversed spin.
The $L$-magnons have lower energy than  the one-magnon excitations up to  a threshold $(|J|/D)_{c1}\simeq 0.14$. 
An overlap  with the two-magnon continuum begins for $(|J|/D)_{c2}\simeq 0.49$. 
We now check how this behavior is reproduced by the approximate approaches.

Mapping to the effective spin-1/2 model, Sec.~\ref{SLM}, yields for $L$-magnons in the anisotropic  ferromagnet 
\begin{equation}
    \label{el_eff_fm}
\varepsilon^{L}_\mk{k} = \varepsilon_0 - \left(\frac{2J^2}{D}-\frac{J^3}{D^2}\right)(1+\gamma_\mk{k}) \ .
\end{equation}
The approximate expression reproduces the exact energy for  the single-ion bound states at ${\bf k}_0 = (\pi,\pi)$.  

The multiboson spin-wave theory  can also be used to compute the
 $L$-magnon dispersion. Similar to the antiferromagnetic case treated in Sec.~\ref{MBSWT},
 the longitudinal magnons have no dispersion in  the harmonic approximation, acquiring dispersion
 due to interactions. A single  cubic
vertex, which exists for a ferromagnet, couples a longitudinal magnon ($c$ boson) with two transverse magnons 
($b$ bosons). Contributions of the quartic terms vanish in the mean-field approximation. We skip straightforward intermediate steps and present only the final result:
\begin{equation}
\label{el_mb_fm_ren}
\varepsilon^c_\mk{k}=\varepsilon_0 +  \frac{(4J)^2}{2N}\sum_\mk{q}
\frac{(\gamma_\mk{q} + \gamma_{\bf k-q})^2}{\varepsilon_0-\varepsilon^b_\mk{q}-\varepsilon^b_{\mk{k}-\mk{q}}}\ .
\end{equation}
The above expression also reproduces the exact value at the Brillouin zone corner $\varepsilon^c_{{\bf k}_0}=\varepsilon_0$.

 In the multiboson theory we can also compute the $L$-magnon decay rate from the imaginary part of the self-energy:
\begin{equation}
\label{Gk_fm}
 \Gamma_\mk{k}=\pi \frac{(4J)^2}{2N} \sum_\mk{q}   
 (\gamma_\mk{q}+\gamma_\mk{k-q})^2\,
 \delta\bigl(\varepsilon_0-{\varepsilon^b_\mk{q}}-{\varepsilon^b_{\mk{k}-\mk{q}}}\bigr).
\end{equation}
This expression vanishes at $\mk{k}_0$ confirming the long-lived character of $L$-magnons at the Brillouin 
zone corner.

Figure \ref{fig_fm_LM_J-D1} presents the two magnon excitation spectra for the anisotropic ferromagnet
along high-symmetry lines in the Brillouin zone. Different panels correspond to  three characteristic values of $J/D$.
Besides the single ion (SIBS) and the exchange bound states (EBS) obtained from the exact solution of 
the two magnon problem, we also show the exact two-magnon continuum and the results for  the $L$-magnons from 
the effective Hamiltonian approach (EH) and by the  multiboson spin-wave theory (MB). For $J/D=-1/5$, panel (a), 
the $L$-magnon branch lies below the two-magnon continuum for all wavevectors.

The exact solution shows that  the binding energy vanishes smoothly at the onset of the two-magnon continuum. 
This recovers  the known property of exponentially-vanishing binding energy of the two-magnon exchange bound states 
in isotropic ferromagnets \cite{Mattis}. The effective spin-1/2 model (EH) is in very good agreement with the exact solution,  
as long as the $L$-magnons are outside the continuum. 

The multiboson theory (MB) also shows a good agreement  with the exact solution for small $|J|/D <  0.5$. 
The deviations are, however, apparent in Figs.~\ref{fig_fm_LM_J-D1}(b) and (c) for larger $|J|/D$, for which 
the longitudinal magnon branch starts to overlap with the continuum. Similar to the antiferromagnet, the anomalous contribution 
to the excitation energy is due to the logarithmic Van Hove singularity in the two-particle density of states, which  transfers 
to the magnon self-energy computed in the Born approximation. To obtain a more physical behavior
one has to resort to  solution of  the Dyson equation keeping a full frequency dependence of $\Sigma_c(\omega,{\bf k})$
\cite{Chernyshev09}. Such calculation is beyond the scope of the present work.

\section{Conclusions}
\label{Concl}

Motivated by experimental observations 
\cite{Fert78,Katsumata00,Legros2021,Bai2021,Psaltakis84,Wyzula22,Mardele24}, 
we  developed a general theory of longitudinal magnons as a distinct type of multipolar excitation
in strongly anisotropic magnetic solids. 
Longitudinal magnons ($L$-magnons) have $S^z=\pm 2S$ and can exist as
coherent, long-lived excitations with large anomalous $g$-factors when $D\agt |J|$. 
Starting with the strong-coupling limit, we have presented two 
alternative analytical methods to compute their dispersion and other properties: the mapping to an effective spin-1/2 
model and the multiboson spin-wave theory.

We compare the results of two approximate methods for the $S=1$ anisotropic square-lattice model (\ref{H}).
For antiferromagnetic couplings $J>0$, the analytical results  are further compared to the numerical linked-cluster
calculations, whereas for the ferromagnetic case $J<0$ we use the exact solution of the two-magnon problem.
Both the effective spin-1/2  model and the multiboson spin-wave theory predict the dispersion
of $L$-magnons with good accuracy up to moderate values of $|J|/D\simeq 0.75$. For  antiferromagnetic coupling, 
the multiboson theory is more accurate of the two approaches. However, in the ferromagnetic state 
the effective spin-1/2 Hamiltonian yields somewhat better results when compared to the exact solution. 

In general, approximations adopted for each of the two analytical approaches used to describe $L$-magnons
can be further improved to obtain more accurate results and to extend their respective ranges of applicability. However,  
a clear advantage of the multiboson spin-wave theory stems from the fact that it treats on equal footing both types of magnons 
and  provides the finite lifetimes of $L$-magnons, when their energies begin to overlap with the multi-magnon continuum.
We have performed such calculations for two-magnon decays of the $L$-magnons in the $S=1$ model 
both for ferromagnetic and antiferromagnetic exchange couplings. The second-order Born approximation leads 
to  singularities both in the real and in the imaginary part of the longitudinal magnon self-energy in line with their general appearance 
for two-dimensional bosonic models  \cite{MZH13}. Such singularities disappear once higher-order renormalization 
processes are taken into account. Finite lifetime calculations enable us to trace the fate of  the multipolar crystal-field excitations 
up to the limit of weak anisotropy, where these excitations completely disappear.

For larger spins $S>1$, spin conservation dictates that the lifetime of 
longitudinal magnons is determined by the $2S$-particle decays. However, in the multiboson 
representation of the spin Hamiltonian there are no bare matrix elements for such processes.  
Instead, an $L$-magnon with the quantum number  $S^z = 2S$ decays into a single magnon with $S^z = 1$
and a composite excitation with $S^z = 2S-1$. The latter decays again with an emission of a single magnon into
an excitation   with $S^z = 2S-2$, etc. Calculation of finite life times due to such cascade multi-particle 
decays is an interesting open problem.

Our results provide a systematic way to describe  longitudinal magnons in various 
materials. In addition to the experimental examples listed in the Introduction, Sec.~\ref{Intro}, 
we also mention here the $S=1$ antiferromagnetic Ni(NCS)$_2$(pyzdo)$_2$ polymer  \cite{Manson23}, which has a 
square lattice geometry resembling the spin model studied in our work.  The moderate value of
 $D/J\alt 1$ suggests that   longitudinal magnons should exist  at finite momenta in certain regions 
 of the Brillouin zone for this material.
 
Yet another very interesting development,  closely related to the scope of our theory, is the recent experimental
observation of the condensation of the single-ion bound states ($L$-magnons)  in a magnetic field applied to 
the spin-1 material Na$_2$BaNi(PO$_4$)$_2$ \cite{Sheng25,Sheng25Oct}.  Bound magnons condensation in 
the saturated phase of this triangular antiferromagnet produces a spin-nematic phase, which persists over a broad range of 
magnetic fields.  The theoretical description of such condensation can be developed by an extension of our theory. 
Possible condensation of longitudinal magnons in anisotropic materials with large spins, as, for example,
spin-2 antiferromagnets FePS$_3$  \cite{Wyzula22} and FePSe$_3$  \cite{Mardele24}
may lead to even more exotic quantum spin phases, with high-rank spin-tensor order parameters. 
 
\section*{Acknowledgement}
We are grateful to D. M. Basko, C. D. Batista, A. L. Chernyshev, R. Coldea,  C. Faugeras, M. Orlita, M. Potemski and 
A. Wildes for valuable  discussions. The financial support was provided in part by the French Research Agency (ANR), 
within the project FRESCO, No.\ ANR-20-CE30-0020. T.Z. would like to thank S. Flach and J.-W. Ryu for hospitality at 
the Center for Theoretical Physics of Complex Systems, Institute for Basic Science, Daejeon, Korea.


\section*{Data Availability}

The data are available from the authors upon reasonable request.

\section*{APPENDIX}

We include here additional details on the multiboson spin-wave theory for the $S=1$ antiferromagnet presented in Sec.~\ref{MBSWT}. 
After the Fourier transformation, the quadratic Hamiltonian for the $b$ bosons acquires the following form:
\begin{equation}
\hat{\cal H}^b_{2} = \sum_{\bf k} \: \Bigl[ A_{\bf k} 
b_{\mk{k}}^\dagger b_\mk{k} - \frac{1}{2} \,
B_{\bf k}\, \bigl( 
b_\mk{k}^\dagger b_{-\mk{k}}^\dagger
+ b_{-\mk{k}} b_\mk{k}
\bigr)\Bigr] ,
\end{equation}
with $A_{\bf k} = \mu$,  $B_{\bf k} = 4J\bar{s}^2\gamma_{\bf k}$,
and $\gamma_{\bf k} = \frac{1}{2}(\cos k_x + \cos k_y)$.
It can be diagonalized  by using the canonical Bogolyubov transformation:
\begin{equation}
\label{Bogolyubov-b}
b_\mk{k}=u_\mk{k}{\tilde{b}}_\mk{k}+v_{\mk{k}}{\tilde{b}}_{-\mk{k}}^\dagger,
\end{equation}
where $u_\mk{k}$ and $v_\mk{k}$ are real and satisfy 
$u_\mk{k}^2-v_\mk{k}^2=1$. The condition on vanishing anomalous terms yields
\begin{equation}\label{u-v}
u_\mk{k}^2 + v_\mk{k}^2 = 
\frac{A_{\bf k}}{\varepsilon_{\bf k}}\ , \quad
2u_\mk{k}v_\mk{k} =
\frac{B_{\bf k}}{\varepsilon_{\bf k}}
\end{equation}
with 
\begin{equation}
\varepsilon_{\bf k} = \sqrt{A_{\bf k}^2 - B_{\bf k}^2}
\end{equation}
being the excitation energy for the Bogolyubov 
$\tilde{b}_\mk{k}$ bosons.

\subsection{Cubic Interactions}

The cubic boson terms $\hat{\cal H}_3$ are represented in the momentum  space as 
\begin{eqnarray}\label{H3-V123}
    \hat{\cal H}_3 & =& \frac{1}{2\sqrt{N}}\sum_{\mk{k},\mk{q}}{V^{(1)}_3(\mk{k},\mk{q})(\tilde{b}_\mk{k}\tilde{b}_\mk{q}c_{-\mk{k}-\mk{q}}+{\rm h.c.})}\nonumber\\
    & & +\frac{1}{\sqrt{N}}\sum_{\mk{k},\mk{q}}{V^{(2)}_3(\mk{k},\mk{q})(\tilde{b}_\mk{k}^\dagger\tilde{b}_\mk{q}c_{\mk{k}-\mk{q}}+{\rm h.c.})}\\
    & &+\frac{1}{2\sqrt{N}}\sum_{\mk{k},\mk{q}}{V^{(3)}_3(\mk{k},\mk{q})(c_\mk{k}^\dagger\tilde{b}_\mk{q}\tilde{b}_{\mk{k}-\mk{q}}+{\rm h.c.})}\nonumber
\end{eqnarray}
where h.c.\ stands for the Hermetian conjugation. The three cubic vertices are expressed by
\begin{eqnarray}\label{V123}
V^{(1)}_3(\mk{k},\mk{q}) & = & 
-4J\bar{s} \bigl( \gamma_\mk{k}u_\mk{k}v_\mk{q} + \gamma_{\mk{q}}u_\mk{q}v_\mk{k}\bigr) \,,
\nonumber\\
V^{(2)}_3(\mk{k},\mk{q}) & = & -4J\bar{s}
\bigl(\gamma_\mk{k}v_\mk{k}v_\mk{q} + \gamma_{\mk{q}}u_\mk{k}u_\mk{q}
\bigr)\,,
\\
V^{(3)}_3(\mk{k},\mk{q}) & = & -4J\bar{s} \bigl(\gamma_{\mk{k}-\mk{q}}u_\mk{q}v_{\mk{k}-\mk{q}} + \gamma_{\mk{q}}u_{\mk{k}-\mk{q}}v_\mk{q}\bigr)\,.
\nonumber
\end{eqnarray}
Here $V^{(1)}_3$ represents the vacuum source vertex, while  $V^{(2,3)}_3$ describe scattering processes. 
Treating the cubic terms as a perturbation to the harmonic theory we obtain in the second order the following 
magnon self energies:
\begin{eqnarray}
&& \Sigma_b(\mk{\omega},\mk{k})  =  
\frac{1}{N}\!\sum_\mk{q}\Biggl[\frac{\bigl|V^{(2)}_3(\mk{k},\mk{q})\bigr|^2}{\mk{\omega}
\!-\!{\varepsilon}^b_\mk{q}\!-\!{\varepsilon}^c_{\mk{k}-\mk{q}}\!+ i0} -
\frac{\bigl|V^{(1)}_3(\mk{k},\mk{q})\bigr|^2}{\mk{\omega}\!+ \varepsilon^b_\mk{q}\!+ \varepsilon^c_{\mk{k}+\mk{q}}}\Biggr]\,\!,
\nonumber \\[2mm]
&& \Sigma_c(\mk{\omega},\mk{k}) =  \frac{1}{2N} \!\sum_\mk{q}\Biggl[\frac{\bigl|V^{(3)}_3(\mk{k},\mk{q})\bigr|^2}{\mk{\omega}\! -{\varepsilon}^b_\mk{q}\! -{\varepsilon}^b_{\mk{k}-\mk{q}}\!+i0}
-\frac{\bigl|V^{(1)}_3(\mk{k},\mk{q})\bigr|^2}{\mk{\omega}\!+\varepsilon^b_\mk{q}\!+\varepsilon^b_{\mk{k}+\mk{q}}}\Biggr]
\nonumber \label{sigma_c_H3}\!,
\end{eqnarray}
where $\varepsilon^b_{\bf k}$ and $\varepsilon^c_{\bf k}$ are the harmonic magnon energies. 
In the Born approximation, the energy corrections to the $b$ and $c$ magnons are given by
${\delta\varepsilon_\mk{k}^{b}}=\Sigma_b(\varepsilon^b_{\bf k},\mk{k})$ and ${\delta\varepsilon_\mk{k}^{c}}^\prime=\Sigma_c(\varepsilon^c_{\bf k},\mk{k})$, respectively.
 
For $J$ exceeding the  critical ratio $(J/D)_{c2}\simeq0.75$, the $c$-magnon branch crosses into the two-magnon continuum
and the self-energy $\Sigma_c(\varepsilon^c_{\bf k},\mk{k})$ acquires a finite imaginary
part. The corresponding decay rate of the $c$-magnons is given in the Born approximation by (\ref{Gk-afm}).
In addition, near the crossing point one finds singularities  in $\Sigma_c(\omega,\mk{k})$ that are briefly
discussed in Sec.~\ref{subsec-afm-comp}, see also \cite{MZH06}. Note, that the $b$-magnon self energy never has an imaginary part
on the mass surface $\omega=\varepsilon^b_{\bf k}$.
\vspace*{1mm}

\subsection{Quartic Interactions}

The Hamiltonian $\hat{\mathcal{H}}_4$ is treated in the mean-field approximation with Hartree-Fock decoupling of the boson operators. The subsequent energy contribution is on the same footing as the one obtained from the cubic Hamiltonian $\hat{\mathcal{H}}_3$ treated perturbatively above. Around the harmonic state, the only non-zero average is $\Delta_b$, Eq. (\ref{delta_b_mf_afm}). The decoupling yields a non-diagonal quadratic Hamiltonian in the $c$ bosons. After applying a Fourier transformation, we obtain
\begin{equation}\label{Ham4-MFT}
    \hat{\mathcal{H}}_{4}^{\text{MF}}=-2J\Delta_b\sum_\mk{k}{\gamma_\mk{k}\left(c_\mk{k}c_{-\mk{k}}+\rm h.c.\right)}.
\end{equation}
As the $b$ magnon spectrum is intact in this procedure we only write the total mean-field Hamiltonian in the $c$ bosons
\begin{eqnarray}\label{Ham24-MFT}
    \hat{\mathcal{H}}^{\text{MF}}& = & \sum_\mk{k}\bigg[(4J\bar{s}^2\! +\!\mu\!-\!D)c_\mk{k}^\dagger c_{\mk{k}}\nonumber\\
    & &\qquad-2J\Delta_b\gamma_\mk{k}\left(c_\mk{k}c_{-\mk{k}}+\rm h.c.\right)\bigg]
\end{eqnarray}
where we include the diagonal part that appears in the Fourier transformation of $\hat{\mathcal{H}}_2$. We diagonalize $\hat{\mathcal{H}}^{\text{MF}}$ by means of a canonical Bogolyubov transformation
\begin{equation}
    c_{\bf k} = w_{\bf k} \tilde{c}_{\bf k} + x_{\bf k} \tilde{c}^\dagger_{-\bf k}
\end{equation}
 where $w_{\bf k}$ and $x_{\bf k}$ are defined in a way similar to (\ref{u-v}). The resulting spectrum of $\hat{\mathcal{H}}^{\text{MF}}$ encodes the renormalized energy in the mean-field processing:
 \begin{eqnarray}
     \varepsilon_\mk{k}^c+{\delta\varepsilon_\mk{k}^{c}}^{\prime\prime}=\sqrt{\left(4J\bar{s}^2  +\mu-D\right)^2-\left(4J\Delta_b\gamma_\mk{k}\right)^2}.~~~~~~
 \end{eqnarray}
 In the leading order, the correction ${\delta\varepsilon_\mk{k}^{c}}^{\prime\prime}$ is given by (\ref{Ek-corr4-L}). 
 
 In total, the magnon spectra in the multiboson spin-wave theory are $\varepsilon_\mk{k}^{b}+{\delta\varepsilon_\mk{k}^{b}}^\prime$
for the $T$-magnons and $\varepsilon_\mk{k}^{c}+{\delta\varepsilon_\mk{k}^{c}}^\prime+{\delta\varepsilon_\mk{k}^{c}}^{\prime\prime}$
for the $L$-magnons.


\end{document}